\documentclass[fleqn,usenatbib]{mnras}

\usepackage{newtxtext,newtxmath}

\usepackage{amssymb}

\usepackage[T1]{fontenc}
\usepackage{subcaption}
\DeclareRobustCommand{\VAN}[3]{#2}
\let\VANthebibliography\thebibliography
\def\thebibliography{\DeclareRobustCommand{\VAN}[3]{##3}\VANthebibliography}

\usepackage{siunitx}
\usepackage{soul,xcolor}
\usepackage{color,soul,natbib}
\usepackage{graphicx}	
\usepackage{amsmath}	
\usepackage{longtable}
\usepackage{mathtools}
\usepackage{ulem}
\usepackage{lineno}
\usepackage{indentfirst}

\title[Long-term double synchronization]{Long-term double synchronization in close-in gas giant planets}

\author[Guo et al.]{
Shuaishuai Guo$^{1,2,3}$,
Jianheng Guo$^{1,2,3}$\thanks{E-mail: guojh@ynao.ac.cn},
Jie Su$^{1,2,3}$,
Dongdong Yan$^{1,2,3}$
\\
$^{1}$Yunnan Observatories, Chinese Academy of Sciences, P.O. Box 110, Kunming 650011, People's Republic of China\\
$^{2}$School of Astronomy and Space Science, University of Chinese Academy of Sciences, Beijing, People's Republic of China\\
$^{3}$Key Laboratory for the Structure and Evolution of Celestial Objects, CAS, Kunming 650011, People's Republic of China
}

\date{Accepted XXX. Received YYY; in original form ZZZ}

\pubyear{2024}

\begin{document}


\label{firstpage}
\pagerange{\pageref{firstpage}--\pageref{lastpage}}
\maketitle

\begin{abstract}
Hot Jupiters, orbiting their host stars at extremely close distances, undergo tidal evolution, with some being engulfed by their stars due to angular momentum exchanges induced by tidal forces. However, achieving double synchronization can prolong their survival. Using the MESA stellar evolution code, combined with the magnetic braking model of Matt et al. (2015), we calculate 25,000 models with different metallicity and study how to attain the conditions that trigger the long-term double synchronization. Our results indicate that massive planets orbiting stars with lower convective turnover time are easier to achieve long-term double synchronization. The rotation angular velocity at the equilibrium point ($\Omega_{\mathrm{sta}}$) is almost equal to orbital angular velocity of planet ($\mathrm{n}$) for the majority of the main sequence lifetime if a system has undergone a long-term double synchronization, regardless of their state at this moment. We further compared our results with known parameters of giant planetary systems and found that those systems with larger planetary masses and lower convective turnover time seem to be less sensitive to changes in the tidal quality factor $Q'_{_*}$. We suggest that for systems that fall on the state of $\Omega_{\mathrm{sta}} \approx n$, regardless of their current state, the synchronization will persist for a long time if orbital synchronization occurs at any stage of their evolution. Our results can be applied to estimate whether a system has experienced long-term double synchronization in the past or may experience it in the future.
\end{abstract}

\begin{keywords}
planet-star interactions -- stars: rotation-- stars: low-mass -- planets and satellites: dynamical evolution and stability
\end{keywords}


\twocolumn

\section{Introduction}

In a close-in binary system, tidal interactions between the two stars tend to align their rotational axes, circularize their orbits, and synchronize their rotations \citep{1977A&A....57..383Z,1981A&A....99..126H,2014ARA&A..52..171O}. The importance of tidal effects on hot Jupiters was first highlighted by \citet{1996Sci...274..954R} after the discovery of the first hot Jupiter,  51Pegasi b \citep{1995Natur.378..355M}. The Kepler, CoRoT, TESS and ground-based telescopes have enabled astronomers to have a deeper insight into the exoplanet systems. In the past thirty years, several hundred hot Jupiters have been discovered around main-sequence FGK stars. Generally, when the orbital period is less than 10$\,$days, their orbital eccentricities are very low \citep{2015ARA&A..53..409W}. According to current theories of planet formation, it is difficult to accumulate the necessary material to form close-in planets around their stars \citep{1996Natur.380..606L}. Therefore, a possibility for the formation of close-in exoplanets is that in the early stages of planet formation, hot Jupiters migrate to short orbits due to the drag of the gaseous protoplanetary disk \citep{1996Natur.380..606L,1998ApJ...500..428T,1997Icar..126..261W,2007A&A...463..775P}.

In various orbital configurations, the double synchronous orbits denote the orbital configuration where the rotational period of a star is nearly equal to the orbital period of its orbiting planet. In the case of conservation of total angular momentum, \citet{1980A&A....92..167H} proves that an equilibrium status of co-planarity, circularity and co-rotation can be attained for a binary system. However, the total angular momentum of such the system can not be conserved because the magnetic braking caused by the stellar winds takes the angular momentum out the system \citep{2015ApJ...799L..23M}. \citet{2012PhDT.......473C} suggested that double synchronous orbits may also decay due to magnetic braking. Therefore, only F-type stars with a massive companion can maintain double synchronous states owing to the weak effect of magnetic braking. The loss of stellar angular momentum can be compensated to some extent by the tidal friction between the star and planet. In fact, the angular momentum will be transferred to the star if the orbital angular velocity of the planet is larger than the angular velocity of the star. Thus, the competition between the magnetic braking and tidal friction determines when an orbital configuration of co-rotation can occur \citep{2015A&A...574A..39D}.

It is important to study the occurrence of the double synchronous orbit (or co-rotation) because the presence of quasi-stable states can significantly delay their tidal evolution and extend the life of the planet, which would otherwise cause the planet to fall into its host star \citep{2015A&A...574A..39D}. There is observational evidence suggesting that close-in companion stars around F-type stars may have larger masses compared to those around hosts of lower mass. \citet{2011A&A...525A..68B} proposed an explanation for this, stating that sufficiently massive companions and fast-spinning primary stars can achieve a tidal equilibrium state with synchronized orbital motion and stellar rotation. Due to weaker magnetic braking, F-type stars can maintain fast rotation throughout their main sequence phase, allowing companion stars to survive for a longer period. \citet{2016A&A...589A..55D} also found, by calculating the timescale for orbital decay, that F-type stars may have significantly longer durations of hosting high-mass companions compared to G-type stars, given a certain orbital period. More recently, \citet{2021MNRAS.505.4956B} discovered a short-period brown dwarf system, GPX-1, orbiting a rapidly rotating F-type young star. The authors found that the rotation period of GPX-1 is synchronized with the orbital period of the brown dwarf, and the system is in a pseudo equilibrium state and stable. In F-type stars, the loss of angular momentum due to stellar wind can be counteracted by the tidal torque induced by the close-in planet, preventing significant changes for the stellar rotation  and planetary orbit. This makes it easier to maintain a double synchronization state in a longer duration.

\citet{2020ApJ...889..108A} found that the evolution of the rotation period of solar-like stars varies with different metallicity. Stars with low metallicity tend to rotate faster, due to their thin convective envelopes and weak magnetic braking. F-type stars, like G-type stars with low metallicity, have thin convective envelopes that result in weaker magnetic braking. This suggests that metallicity affects the balance between wind torque and tidal torque, which in turn affects orbital stability. Furthermore, the tidal interactions between the star and planet are also affected by the properties of the planet, such as planetary mass and separation. However, there is lack of sufficient investigation for the co-rotation of planets around late-type stats. It is not clear how to trigger double synchronization around late-type stars with different metallicities. Thus, these findings motivate us to investigate the phenomenon of the double synchronization further. Our aim is to explore the conditions that trigger the double synchronous.

In this paper, we focus on hot Jupiters orbiting G and F types stars. We replaced the Skumanich-type law \citep{1972ApJ...171..565S} with the magnetic braking scheme of \citet{2015ApJ...799L..23M} and apply the criterion of \citet{2015A&A...574A..39D} to determine if the star-planet system is in the state of co-rotation. We briefly described in section \ref{subsec:interaction} the grid spacing we used in our MESA calculations. We reviewed previous studies in Section \ref{subsec:Pseudo-stable} and described the conditions for pseudo-stable equilibrium of hot Jupiters given by \citet{2015A&A...574A..39D}. In Section \ref{subsec:Balance}, we discussed the stellar spin rate $\Omega_{\text{sta}}$ at tidal and wind torque balance based on the magnetic braking model of \citet{2015ApJ...799L..23M}. In Section \ref{subsec:metallicity}, we examined the impact of stellar mass, metallicity, and initial spin rate on triggering double synchronization. In Section \ref{subsec:planet mass}, we discussed the impact of planetary mass, tidal quality factor, and initial semi-major axis on triggering double synchronization. In Section \ref{subsec:Synchronicity}, we quantitatively present the distribution of initial parameters for the occurrence of long-term double synchronization and discuss them. In Section \ref{sec:disscussion}, we have explored the distribution of 243 observational samples within the Darwinian diagram and the phases during which long-term double synchronization occurs.

\section{Method and Model} \label{sec:Model}
\subsection{Star-planet interaction model in MESA} \label{subsec:interaction}
In our previous work, we incorporated the process of angular momentum exchange between stars and planets into the stellar evolution code MESA, version 11554 \citep{2023RAA....23i5014G}. We did not consider the differential rotation of the star, which is reasonable. Taking the Sun as an example, the error between considering and not considering the differential rotation does not exceed 5\%. tectbf{The} detailed calculation procedures for this are provided in Appendix \ref{subsec:sun}. MESA has been widely utilized in the field of stellar and planetary evolution \citep{2011ApJS..192....3P,2013ApJS..208....4P,2015ApJS..220...15P,2018ApJS..234...34P,2019ApJS..243...10P}. The values of the parameters we adopted in this study are summarized in Table \ref{tab:tab1}. It is worth noting that the initial parameters for the nearly 25,000 models we calculated are all possible combinations of the six parameters listed in Table \ref{tab:tab1}.

\begin{table}
\begin{center}
\caption{ All possible initial parameter combination.} \label{tab:tab1}
 \begin{tabular}{clclclcl}
  \hline\noalign{\smallskip}
Parameter &  &  &  &  &  &  &                    \\
  \hline\noalign{\smallskip}
$M_{*}$$\,$($M_{\odot}$) & 0.8 & 0.9 & 1.0 & 1.1 & 1.2 & 1.3 & --  \\
$[Fe/H]$$\,$(dex) & -0.5 & -0.3 & 0.0  & +0.3 & +0.5 & -- & -- \\
$M_{\mathrm{p}}$$\,$($M_{\mathrm{J}}$)& 0.1 & 0.5 & 1.0 & 4.0 & 8.0 & 13.0 & --   \\
$P_{\mathrm{rot, ini}}$$\,$(days) & 0.6 & 3.0 & 8.0 & 16.0 & -- & -- & --   \\
$Q'_{_*}$ & $10^{5}$ & $10^{6}$ & $10^{7}$ & $10^{8}$ & $10^{9}$ &-- &--  \\
$a_{\mathrm{ini}}$$\,$(au) & 0.01 & 0.02 & 0.03 & 0.04 & 0.05 & 0.06 & 0.10   \\
  \noalign{\smallskip}\hline
\end{tabular}
\end{center}
\end{table}

\subsection{Pseudo-stable equilibrium of hot Jupiters} \label{subsec:Pseudo-stable}
In this section we summarize \citet{2015A&A...574A..39D} results. \citet{1980A&A....92..167H} found that in binary systems, when the total angular momentum of the system exceeds a critical value, i.e. $L > L_{\mathrm{c_0}}$, the system will reach a double synchronous rotation state, which can prevent the planet from further inward migration. Subsequently, \citet{2015A&A...574A..39D} found the conditions for maintaining a pseudostable equilibrium state when taking into account the stellar magnetic braking. They derived the conditions for tidal pseudo-equilibrium under magnetic braking and showed that the system evolves towards a circular and aligned orbit with minimum energy. The pseudo-stable condition is given by:
\begin{equation}
\textmd{n} = \sqrt{\frac{G(M_*+M_{\mathrm{pl}})}{{a}^3}}
\label{eq1}
\end{equation}
\begin{equation}
\omega = \textmd{n}
\label{eq2}
\end{equation}
\begin{equation}
\Omega = \beta \textmd{n}
\label{eq3}
\end{equation}
where $\omega$ and $\Omega$ are the angular velocity of the planet and the star, respectively. $a$ is the semi-major axis, $M_{*}$ and $M_{\mathrm{pl}}$ are the stellar and planetary masses. n is the mean orbital motion, and $\beta$ is a function of time defined as:
\begin{equation}
\beta = 1 - \frac{d\mathrm{L}}{d\mathrm{t}}(I_*\frac{\mathrm{d}\Omega}{\mathrm{d}t})^{-1}
\label{eq4}
\end{equation}
Due to the presence of magnetic braking, the total angular momentum of the system gradually decreases, i.e. $\frac{\mathrm{dL}}{\mathrm{dt}} < 0$. When $\beta = 1$, the system has $\Omega$ = n, and the star and planet are in double synchronization. At this point, $\frac{\mathrm{d}\Omega}{\mathrm{d}t} \approx 0$, and the total angular momentum of the system is approximately conserved, equivalent to the case without considering magnetic braking. When the tidal torque is negligible compared to magnetic breaking, $\beta \approx 0$. When $0 < \beta < 1$, the system has $\Omega < n$, and $\frac{\mathrm{d}\Omega}{\mathrm{d}t} < 0$ is required for the equation to hold, i.e. the tidal torque is less than the wind torque. When $\beta > 1$, the planet migrates outward under tidal action, and both tidal torque and wind torque slow down the stellar rotation. However, the equation requires $\frac{\mathrm{d}\Omega}{\mathrm{dt}} > 0$, which is clearly not true. Therefore, only when $\beta < 1$, the equilibrium state can be reached. Moreover, the equilibrium is stable only if the total angular momentum of the system $L > L_{\mathrm{c}}$ and $L_{\mathrm{orb}} > L_{\mathrm{orb,s}}$, where

\begin{equation}
L_{\mathrm{orb}} = \frac{M_{*}M_{\mathrm{pl}}}{M_{*}+M_{\mathrm{pl}}}na^2
\label{eq5}
\end{equation}
\begin{equation}
L_{\mathrm{orb,s}} = (4-\beta)(I_{*}+I_{\mathrm{sp}})n
\label{eq6}
\end{equation}
\begin{equation}
L = G^{2/3}\frac{M_{*}M_{\mathrm{pl}}}{(M_{*}+M_{\mathrm{pl}})^{1/3}}n^{-1/3}+(\beta I_{*}+I_{\mathrm{sp}})n
\label{eq7}
\end{equation}
\begin{equation}
L_{\mathrm{c}} = 4\left[\frac{G^2}{3^3}\frac{M_{*}^3M_{\mathrm{pl}}^3}{M_{*}+M_{\mathrm{pl}}}(\beta I_{*}+I_{\mathrm{sp}})\right]^{1/4}
\label{eq8}
\end{equation}
where $L$, $L_{\mathrm{c}}$, $L_{\mathrm{orb}}$ and $L_{\mathrm{orb,s}}$ are the total angular momentum, critical total angular momentum, orbital angular momentum and critical orbital angular momentum respectively. $I_*$ and $I_{\mathrm{sp}}$ are the stellar and planetary moment of inertia.
When $\beta$ = 1, the equation degenerates to the case without magnetic braking. In this case, we have:
\begin{equation}
L_{\mathrm{orb,c_0}} = 3(I_{*}+I_{\mathrm{sp}})n
\label{eq9}
\end{equation}
\begin{equation}
L_{c_0} = 4\left[\frac{G^2}{3^3}\frac{M_{*}^3M_{\mathrm{pl}}^3}{M_{*}+M_{\mathrm{pl}}}(I_{*}+I_{\mathrm{sp}})\right]^{1/4}
\label{eq10}
\end{equation}
In the presence of magnetic braking, it can be demonstrated that $n_{\mathrm{s}}$, defined as the maximum orbital frequency allowing for the existence of pseudo-stable states, is determined by:
\begin{equation}
n_{\mathrm{s}} = \left[G^2\frac{M_{*}^3M_{\mathrm{pl}}^3}{M_{*}+M_{\mathrm{pl}}}\right]^{1/4}(4-\beta)^{-3/4}(I_{*}+I_{\mathrm{sp}})^{-3/4}
\label{eq11}
\end{equation}
At $L = L_{\mathrm{c_0}}$, it corresponds to the only average motion of rotation
\begin{equation}
n_{\mathrm{c_0}} = \left[\frac{G^2}{3^3}\frac{M_{*}^3M_{\mathrm{pl}}^3}{M_{*}+M_{\mathrm{pl}}}\right]^{1/4}(I_{*}+I_{\mathrm{sp}})^{-3/4}
\label{eq12}
\end{equation}
At $n = n_{\mathrm{s}}$ the total angular momentum is
\begin{equation}
L_{s} = \left[G^2\frac{M_{*}^3M_{\mathrm{pl}}^3}{M_{*}+M_{\mathrm{pl}}}(4-\beta)(I_{*}+I_{\mathrm{sp}})\right]^{1/4}\left(1+\frac{\beta I_{*}+I_{\mathrm{sp}}}{(4-\beta)(I_{*}+I_{\mathrm{sp}})} \right)
\label{eq13}
\end{equation}
When $0 < \beta < 1$, we have
$\frac{3}{4}^{\frac{3}{4}}n_{\mathrm{c_0}} < n_{\mathrm{s}} < n_{\mathrm{c_0}}$.
The corresponding conditions for the existence of a pseudo-stable state under total angular momentum can be demonstrated to satisfy $L_{\mathrm{c}} < L_{\mathrm{s}} < L_{\mathrm{c_0}}$. Therefore, as long as the mean motion of the orbit is $n < \frac{3}{4}^{\frac{3}{4}}n_{\mathrm{c_0}}$ and $L > L_{\mathrm{c_0}}$, the system can achieve a pseudo-stable equilibrium.
The Roche limit is defined as
\begin{equation}
a_{\mathrm{R}} = 2.422R_{\mathrm{p}}(\frac{M_{*}}{M_{\mathrm{pl}}})^{1/3}
\label{eq14}
\end{equation}
where $R_{\mathrm{p}} = 1.3$\,$R_{\mathrm{J}}$ and the rotational angular momentum of the planet is negligible compared to its orbital angular momentum, we can neglect $I_{\mathrm{sp}}$.

\subsection{Balance of tidal torque and wind torque} \label{subsec:Balance}

In this section, we did not use the Skumanich-type law \citep{1972ApJ...171..565S}, which assumes that the magnitude of the torque due to magnetic braking is $\dot{\omega_{\mathrm{mb}}} = \alpha_{\mathrm{mb}}\Omega^3$, where $\alpha_{\mathrm{mb}}$ = 1.5 $\times$ $10^{-14}$ $\gamma$ yr and $\gamma$ = 1.0 for G stars and $\gamma$ = 0.1 for F stars. Instead, we followed the work of \citet{2015ApJ...799L..23M} and used their braking equation to find the equilibrium position $\Omega_{\mathrm{sta}}$ of the star, where the tidal and wind torques balance. This equation takes into account not only a simple constant multiplied by $\Omega^3$, but also the thickness of the convective envelope, which determines the magnetic braking resulting from the stellar mass and metallicity. This approach allows us to more accurately reflect the relationship between star mass and magnetic braking, rather than simply distinguishing F and G stars based on their $\gamma$ values. We can also discuss the equilibrium position $\Omega_{\mathrm{sta}}$ of stars with different metallicity. $\Omega_{\text{sta}}$ represents the physically meaningful solution when the torque is balanced, i.e., when $\mathrm{d}\Omega/\mathrm{d}t = 0$. Following \citet{1963MNRAS.126..257G,1968aitp.book.....K,2008ApJ...678.1396J}, the time evolution of the semi-major axis can be expressed as

\begin{equation}
\frac{\mathrm{da}}{\mathrm{dt}} = -\frac{9}{2} \sqrt{\frac{G}{aM_*}} (\frac{R_*}{a})^5 \frac{M_{\mathrm{pl}}}{Q'_{_*}}\left(\frac{{n}-{\Omega}}{|{n}-{\Omega}|}\right)
\label{eq15}
\end{equation}

The total angular momentum of the system:
\begin{equation}
\frac{\mathrm{d}L_{*}}{\mathrm{d}t}=\frac{\mathrm{d}L_{wind}}{\mathrm{d}t}-\frac{\mathrm{d}L_{orb}}{\mathrm{d}t}
\label{eq16}
\end{equation}
Here, $L_{*}$, $L_{\mathrm{orb}}$ and $L_{\mathrm{wind}}$ are the stellar angular momentum, planetary orbital  angular momentum, and the angular momentum lost by the magnetic stellar wind, respectively. They are given by:

\begin{equation}
\frac{\mathrm{d}L_{*}}{\mathrm{d}t}=I_* \frac{\mathrm{d}\Omega}{\mathrm{d}t} + \Omega_{*}\frac{\mathrm{d}{I_*}}{\mathrm{d}t}
\label{eq17}
\end{equation}

\begin{equation}
\frac{\mathrm{d}L_{orb}}{\mathrm{d}t}=\frac{1}{2}\frac{M_{*}M_{\mathrm{pl}}}{M_{*}+M_{\mathrm{pl}}}na\frac{\mathrm{da}}{\mathrm{dt}}
\label{eq18}
\end{equation}

\begin{equation}
\frac{\mathrm{d}L_{wind}}{\mathrm{d}t}=-T_0(\frac{\tau_{\text{cz}}}{\tau_{cz\odot}})^p(\frac{\Omega}{\Omega_\odot})^{p+1} \quad (\text{unsaturated})
\label{eq19}
\end{equation}

\begin{equation}
\frac{\mathrm{d}L_{wind}}{\mathrm{d}t}=-T_0{\chi}^p(\frac{\Omega}{\Omega_\odot}) \quad (\text{saturated})
\label{eq20}
\end{equation}
Due to the relatively small variation in stellar moment of inertia during the main sequence for solar-like stars, the second term on the right side of Equation (\ref{eq17}) is much smaller than the first term. For simplicity, we neglect the second term in Equation (\ref{eq17}). By substituting Equation (\ref{eq15}) into Equation (\ref{eq18}) and utilizing the above equations, we can solve for Equation (\ref{eq21}) and Equation (\ref{eq22}). The negative sign on the left side of the equations represents the tidal effects, while the negative sign on the right side represents the influence of magnetic braking.

\begin{equation}
\begin{split}
\frac{\mathrm{d{\Omega}}}{\mathrm{d{t}}} = \frac{1}{I_*}\left(\frac{9}{4{Q'_{_*}}}\frac{{M_*}^\frac{1}{2}{M_{\mathrm{pl}}}^2{R_*}^5}{G(M_*+M_{\mathrm{pl}})^\frac{5}{2}}\left(\frac{{n}-{\Omega}}{|{n}-{\Omega}|}\right){n}^4 - T_{\mathrm{0}}(\frac{\tau_{\mathrm{cz}}}{\tau_{\mathrm{cz\odot}}})^p(\frac{\Omega}{\Omega_\odot})^{p+1}\right)        
\label{eq21}
\end{split}
\end{equation}
(unsaturated)

\begin{equation}
\frac{\mathrm{d{\Omega}}}{\mathrm{d{t}}} = \frac{1}{I_*}\left(\frac{9}{4{Q'_{_*}}}\frac{{M_*}^\frac{1}{2}{M_{\mathrm{pl}}}^2{R_*}^5}{G(M_*+M_{\mathrm{pl}})^\frac{5}{2}}\left(\frac{{n}-{\Omega}}{|{n}-{\Omega}|}\right){n}^4 - T_{\mathrm{0}}{\chi}^p(\frac{\Omega}{\Omega_\odot})\right)      (saturated)
\label{eq22}
\end{equation}

\begin{equation}
T_{\mathrm{0}} = K(\frac{R_{*}}{R_\odot})^{3.1}(\frac{M_{*}}{M_\odot})^{0.5}{\gamma}^{2m}
\label{eq23}
\end{equation}

\begin{equation}
\gamma = \sqrt{1+(u/0.072)^2}
\label{eq24}
\end{equation}
\begin{equation}
u = \frac{\Omega}{\Omega_{\mathrm{crit}}},\Omega_{\mathrm{crit}} = \sqrt{\frac{GM_*}{{R_*}^3}}
\label{eq25}
\end{equation}
where $R_*$, $\Omega$, and $\Omega_{\mathrm{crit}}$ are the stellar radius, angular rotation rate, and critical angular rotation rate, respectively. The constants $K$, $m$, $p$, and $\chi$ are free parameters. The saturated and unsaturated states in Equations (\ref{eq19}) and (\ref{eq20}) represent two different states of stellar magnetic activity, both of which have a strong correlation with the Rossby number $R_{\mathrm{o}} = \frac{2\pi}{\Omega\tau_{\mathrm{cz}}}$. The convective turnover time $\tau_{\mathrm{cz}}$ can be expressed as \citep{2021ApJ...912...65G}:
\begin{equation}
\tau_{\mathrm{cz}}(r) = \alpha_{\mathrm{MLT}}H_{\mathrm{P}}(r)/v_{\mathrm{c}}(r)
\label{eq26}
\end{equation}
where $H_{\mathrm{P}}(r)$ is the scale height, $v_{\mathrm{c}}(r)$ is the convective velocity at radius $r$, and $\alpha_{\mathrm{MLT}}$ is the convective mixing length. In Equation (\ref{eq26}), $\tau_{\mathrm{cz}}$ is the turnover timescale of the convective zone, which is defined in the location where $r$ = $r_{\mathrm{BCZ}}$ + $0.5$\,$H_{\mathrm{P}}(r)$, and $r_{\mathrm{BCZ}}$ is the radius of the bottom of the outer convection zone. The values of $\alpha_{\mathrm{MLT}}$ is 1.82.

$R_{\mathrm{o}}$ $\leqslant$ $R_{\mathrm{osat}}$ is the saturation region, and vice versa. The parameter $\chi$ = $R_{\mathrm{o}}$ /$R_{\mathrm{osat}}$ relies on the critical value $R_{\mathrm{osat}} = 0.14$ \citep{2018MNRAS.479.2351W}. The solar $R_{\mathrm{o}}$ is around 2 \citep{2016MNRAS.462.4442S}. Thus we take $\chi$ = 14. The constant $m$ is set to be 0.22. In order to reproduce the current rotation period of the Sun, the values of $K$ and $p$ are 1.2 $\times$ $10^{30}$ $erg$ and 2.6, respectively.

Here, the equilibrium point where the tidal and wind torques are balanced corresponds to d$\Omega$/dt = 0. By solving a high-degree one-dimensional function, we can obtain $\Omega_{\text{sta}}$ at the equilibrium point. However, unlike the cubic function in \citet{2015A&A...574A..39D} that can be solved using Cardano's method, we can only seek numerical solutions for high-degree one-dimensional functions. We provide a method for finding the roots of high-degree one-dimensional functions in Appendix \ref{subsec:py}.

\section{Results} \label{sec:result}
\subsection{The influence of stellar parameters on the triggering of double synchronization} \label{subsec:metallicity}

In Figure \ref{fig:1.1M}, we discuss the effects of different initial stellar rotation periods, metallicity, and stellar masses of the star-planet systems on the triggering of the double synchronization mechanism. In Figure \ref{fig:1.1M} (a), we show the evolution of the stellar rotation period and the planetary orbital period, the ratio of orbital angular momentum to critical orbital angular momentum for the system with $M_{*}=1.1$\,$M_{\odot}$, $M_{\text{pl}}=8.0$\,$M_{\text{J}}$ and $a_{\text{ini}}$=0.05$\,$au when the value of $Q'{_*}$ is $10^6$. We also show the ratio of total angular momentum to critical total angular momentum as a function of age. For top panel: the solid colored lines represent the stellar rotation periods $P_{\text{rot}}$, the dashed colored lines represent the planet orbital periods $P_{\mathrm{orb}}$, and the thin solid black line represent the Roche limit. Middle panel: the solid colored lines represents $\frac{L_{\mathrm{orb}}}{L_{\mathrm{orb,c_0}}}$, and the thin dashed black line represents the $L_{\text{orb}} = L_{\mathrm{orb,c_0}}$. In the bottom panel, the solid colored lines represent $\frac{L}{L_{\mathrm{c_0}}}$, and the thin dashed black line represents $L = L_{\mathrm{c_0}}$. In Figure \ref{fig:1.1M} (b), we present the Darwin diagram. The solid colored lines represent the locus of the star-planet system for different metallicities and initial stellar rotation periods. The dashed and solid black lines represent the locus for $\Omega = n$ and $\Omega = 0$, respectively. The colored dash-dotted lines represent the locus when the rate of magnetic braking angular momentum loss is equal to the rate of planetary orbital transfer angular momentum to the star ($\dot{\Omega}$ = 0). The vertical solid black line represents the position of the Roche limit. Thin horizontal dashed black line and thin vertical dashed black line indicate the values $L = L_{\mathrm{c_0}}$ and $\frac{n}{n_{\mathrm{c_{0}}}} = \frac{3}{4}^{\frac{3}{4}}$, respectively. The square markers represent the 99$\%$ position of the system's evolutionary timescale, while the pentagon markers indicate the end point of the system's evolution. The magnified inset is included to provide a clearer view for the example with [Fe/H] = -0.5$\,$dex and $P_{\mathrm{rot,ini}} = 3\,days$. In this scenario, the planetary system, due to the double synchronization effect, can survive until the end of the stellar main-sequence phase.

The top panel of Figure \ref{fig:1.1M} (a) shows that long-term double synchronization is possible when [Fe/H] = -0.5$\,$dex, because the stellar magnetic braking is weak. For example, for the cases with initial rotation periods of 3.0$\,$days and 8.0$\,$days, long-term double synchronization can be maintained for a long time when the stellar rotation rate synchronizes with the planetary orbital rate. However, for [Fe/H] = 0$\,$dex and +0.5$\,$dex, for the case with initial rotation period of 8.0$\,$days, the stellar rotation period and the planetary orbital period approach synchronization only when the planet is about to be engulfed, and for the case with initial rotation period of 3.0$\,$days, due to the effect of magnetic braking in the early stage of evolution and the transfer of stellar rotation angular momentum to the planetary orbit, the stellar rotation period rapidly increases until the stellar rotation rate and the planetary orbital rate synchronize, but this synchronized state is immediately broken. This is because in this case, the stars with [Fe/H] = 0$\,$dex and +0.5$\,$dex have stronger magnetic braking than the stars with [Fe/H] = -0.5$\,$dex, and long-term double synchronization is difficult to maintain. Moreover, long-term double synchronization requires that $\Omega_{\mathrm{sta}} \approx n = \Omega $, which can also be seen in Figure \ref{fig:1.1M} (b), where the systems with [Fe/H] = 0 dex and +0.5$\,$dex do not satisfy $\Omega_{\mathrm{sta}} < n$(red and blue dash-dotted lines), long-term double synchronization cannot be maintained, while the system with [Fe/H] = -0.5$\,$dex satisfies $\Omega_{\mathrm{sta}} \approx n$(green dash-dotted line), long-term double synchronization can be maintained. It is worth noting that from the middle and bottom panels of Figure \ref{fig:1.1M} (a), all models undergo orbital decay and magnetic braking, the $\frac{L_{\mathrm{orb}}}{L_{\mathrm{orb,c_0}}}$ and $\frac{L}{L_{\mathrm{c_0}}}$ ratios of the system with initial stellar rotation periods of 3.0$\,$days and [Fe/H] = -0.5$\,$dex always remain above 1, while the other curves will eventually lead to $\frac{L_{\mathrm{orb}}}{L_{\mathrm{orb,c_0}}}$ and $\frac{L}{L_{\mathrm{c_0}}}$ ratios below 1 as they evolve. When $\frac{L_{\mathrm{orb}}}{L_{\mathrm{orb,c_0}}} < 1$, the planets fall into the star quickly and are swallowed. This is because the total angular momentum of the system is always decreasing due to the effect of stellar magnetic braking, and the higher the stellar metallicity and the thicker the convective envelope, the stronger the magnetic braking, so that the system is unstable when $\frac{L_{\mathrm{orb}}}{L_{\mathrm{orb,c_0}}}$ and $\frac{L}{L_{\mathrm{c_0}}}$ are less than 1. For [Fe/H] = -0.5$\,$dex, we find the stellar-planet system with an initial period of 3.0$\,$days has a larger initial angular momentum, the angular momentum loss caused by magnetic braking is insufficient to decrease the total angular momentum below the critical value during the main sequence phase of the star from the bottom panel of Figure \ref{fig:1.1M} (a).

Meanwhile, from the Darwin diagram in Figure \ref{fig:1.1M}(b), for [Fe/H] = -0.5$\,$dex and $P_{\text{rot,ini}}=3.0$\,$days$, when $\beta > 1$, the planet migrates outward, resulting in a decrease in the orbital velocity during the evolution. Both tidal torque and wind torque contribute to the reduction of the stellar angular momentum, until the system reaches double synchronization at $\beta = 1$. On the other hand, for [Fe/H] = -0.5$\,$dex and $P_{\text{rot,ini}}=8.0$\,$days$, when $0 < \beta < 1$, the tidal torque causes the planet to migrate inward, leading to an increase in the orbital velocity during the evolution. Initially, the tidal torque dominates the wind torque, and the angular momentum remains approximately constant, staying nearly horizontal during the early stages of evolution. However, when the wind torque and tidal torque become comparable, the trajectory follows $\Omega = \Omega_{\mathrm{sta}}$. It is worth noting that the curve at $\dot{\Omega}$ = 0 almost coincides with $\Omega = n$ when [Fe/H] = -0.5$\,$dex. Therefore, as long as the wind torque remains comparable to the tidal torque, the evolution will proceed at almost constant rotation frequency. This can explain why the purple and green solid lines can evolve with almost constant stellar rotation and orbital frequency when they reach double synchronization until stability conditions are destroyed. The purple curve gradually crosses the equilibrium point at $\frac{n}{n_{\mathrm{c_{0}}}} > \frac{3}{4}^{\frac{3}{4}}$, where tidal torque becomes greater than wind torque, becoming unstable and eventually leading to the planet being consumed. When [Fe/H] = 0$\,$dex and [Fe/H] = +0.5$\,$dex, the stars undergo stellar contraction during the PMS phase, resulting in a gradual decrease in radius. This leads to a reduction in the stellar moment of inertia $I_{*}$. According to equations \ref{eq10} and \ref{eq12}, it can be understood that $L_{\mathrm{c_0}}$ decreases and $n_{\mathrm{c_0}}$ increases. Therefore, the evolutionary trajectory shows a trend towards the upper left direction. For the case of [Fe/H] = 0$\,$dex (red and light blue solid lines), as evolution proceeds, the increasing tidal torque approaches the decreasing wind torque, and evolution will follow the trajectory $\Omega = \Omega_{\mathrm{sta}}$ for a period of time, until the planet quickly falls into the star when $\frac{n}{n_{\mathrm{c_{0}}}} > 1$. For the case of [Fe/H] = +0.5$\,$dex (blue and yellow solid lines), 99$\%$ of the planetary life is spent in the saturated phase of the star, and the equilibrium point (blue dash-dotted line) almost coincides with $\dot{\Omega}$ = 0. Due to the thicker convection envelope, magnetic braking is stronger, and the wind torque dominates, causing the star to rotate slowly. As evolution proceeds until the star is in an unsaturated state, the system evolves below the blue dash-dotted line, at which point the tidal torque dominates, causing the planet to be rapidly consumed and the star to spin up.

The Figure \ref{fig:1.3M} is similar to Figure \ref{fig:1.1M}, and the only difference is that the stellar mass is 1.3$\,$$M_{\odot}$. For a 1.3$\,$$M_{\odot}$ star, both planetary systems with [Fe/H] = -0.5$\,$dex and [Fe/H] = 0$\,$dex can trigger the mechanism of long-term double synchronization. Similar to the case of 1.1$\,$$M_{\odot}$, in the system with an initial rotation period of 3.0$\,$days and [Fe/H] = -0.5$\,$dex, the planets can still survive until the end of the stellar main sequence. In the Darwin diagram of Figure \ref{fig:1.3M}(b), we can see that the $\dot{\Omega}$ = 0 equilibrium points for [Fe/H] = -0.5$\,$dex and [Fe/H] = 0$\,$dex almost coincide with $\Omega = n$. When $\beta > 1$ (green and red solid lines), the planet migrates outward, and $n/n_{\mathrm{c_0}}$ decreases. When $\Omega = n$, green and red solid lines is located exactly at $\Omega = \Omega_{\mathrm{sta}}$ in the Figure \ref{fig:1.3M}(b), and the torque from the wind is equal to the tidal torque. The evolution will follow the trajectory of $\Omega = \Omega_{\mathrm{sta}}$. For the case of [Fe/H] = 0$\,$dex, when $\frac{n}{n_{\mathrm{c_{0}}}} > \frac{3}{4}^{\frac{3}{4}}$, the planet gradually crosses the equilibrium point, and the tidal torque is greater than the wind torque. The evolution is below the red dash-dotted line, and the planet quickly is engulfed by the star. When $0 < \beta < 1$ (purple and light blue solid lines), the planet migrates inward, and because the tidal torque is greater than the wind torque, the evolution proceeds almost horizontally until $\Omega = n$. The evolution then follows the trajectory of $\Omega = \Omega_{\mathrm{sta}}$ and gradually crosses the equilibrium point as $\frac{n}{n_{\mathrm{c_{0}}}} > \frac{3}{4}^{\frac{3}{4}}$, eventually being engulfed by the star. For [Fe/H] = +0.5$\,$dex, the evolution is the same as in the case of 1.1$\,$$M_{\odot}$ during the PMS phase. The stellar contraction will cause $L_{\mathrm{c_0}}$ to decrease and $n_{\mathrm{c_0}}$ to increase, and the evolution will proceed in the upper left direction. Until the tidal torque dominates (blue dash-dotted line below), the planet will quickly fall into the star.
\begin{figure*}
    \centering
    \begin{subfigure}[t]{0.99\textwidth}
           \centering
           \includegraphics[width=0.9\textwidth]{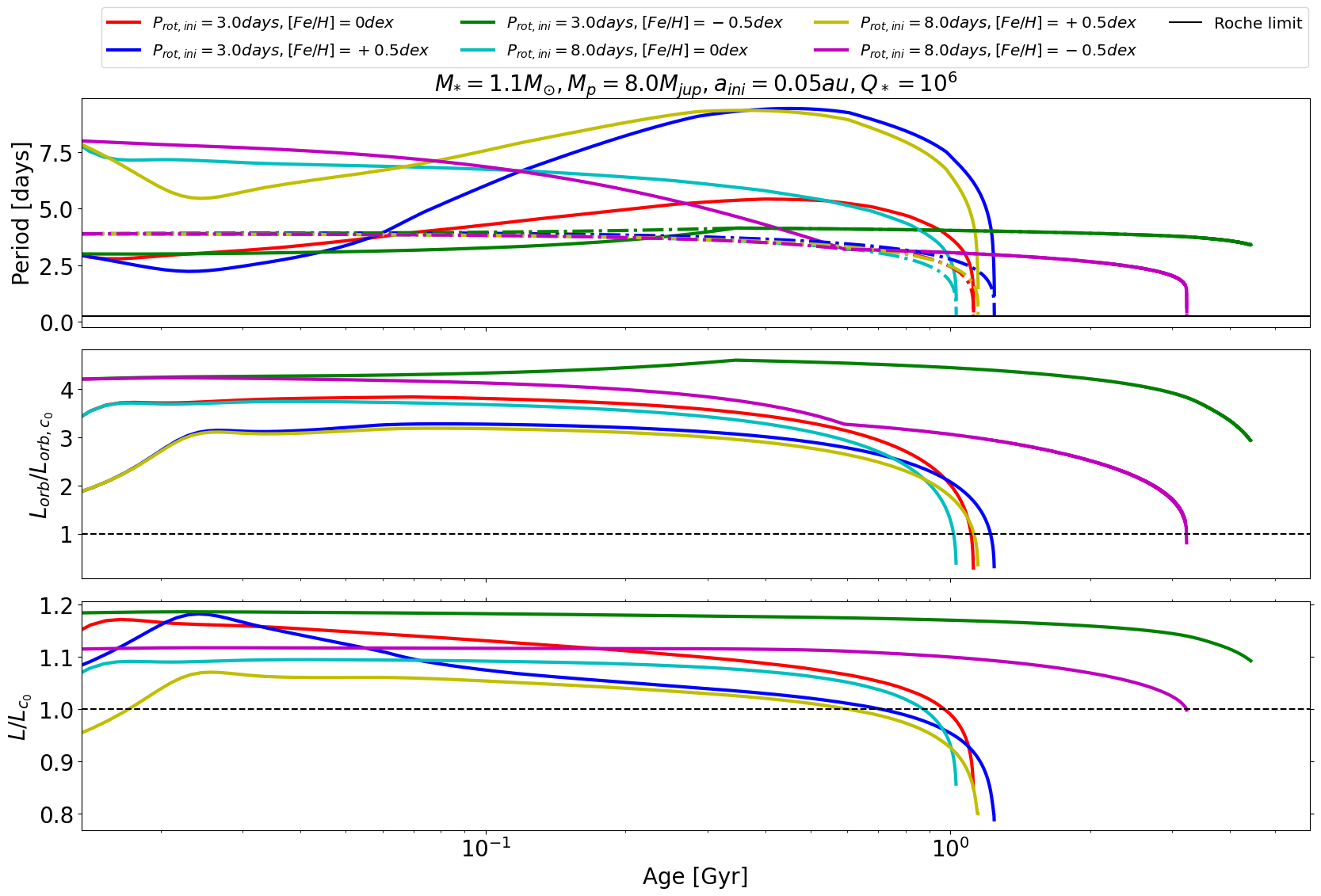}
            \caption{}
            \label{fig1:a}
    \end{subfigure}
    \begin{subfigure}[t]{0.99\textwidth} \ContinuedFloat
            \centering
            \includegraphics[width=0.9\textwidth]{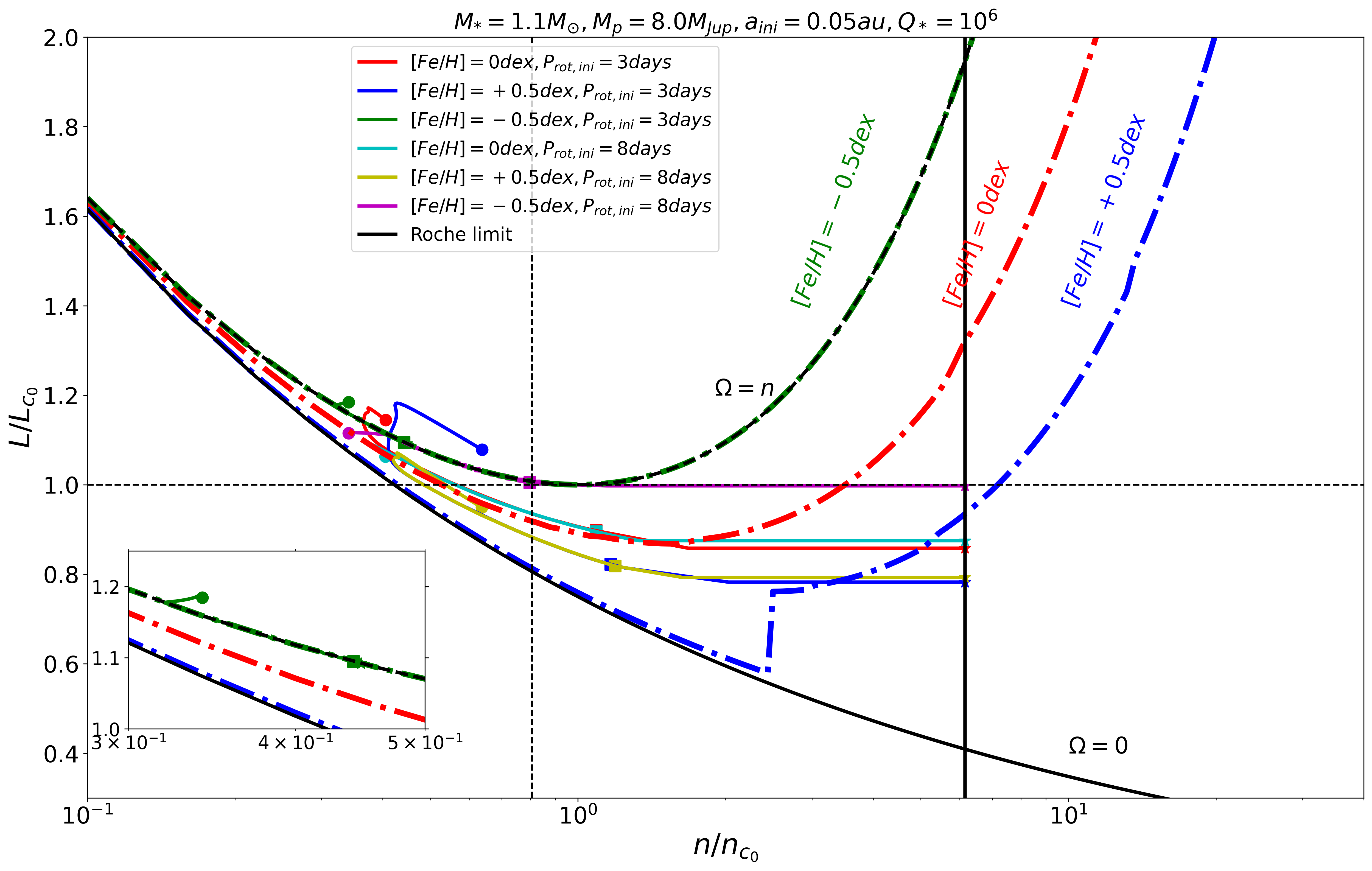}
            \caption{}
            \label{fig1:b}
    \end{subfigure}
\caption{The panel (a) shows the relationship between age and various parameters, including stellar rotation period, orbital period, ratio of orbital angular momentum to critical orbital angular momentum, and ratio of total angular momentum to critical total angular momentum.  The panel (b) shows the Darwin diagram of the evolution of planetary system angular momentum under the effects of tidal dissipation and magnetic braking. (The specific meaning of the curves in the figure can be found in the first paragraph of Section \ref{subsec:metallicity}.)
\label{fig:1.1M}}
\end{figure*}

\begin{figure*}
    \centering
    \begin{subfigure}[t]{0.99\textwidth}
           \centering
           \includegraphics[width=0.9\textwidth]{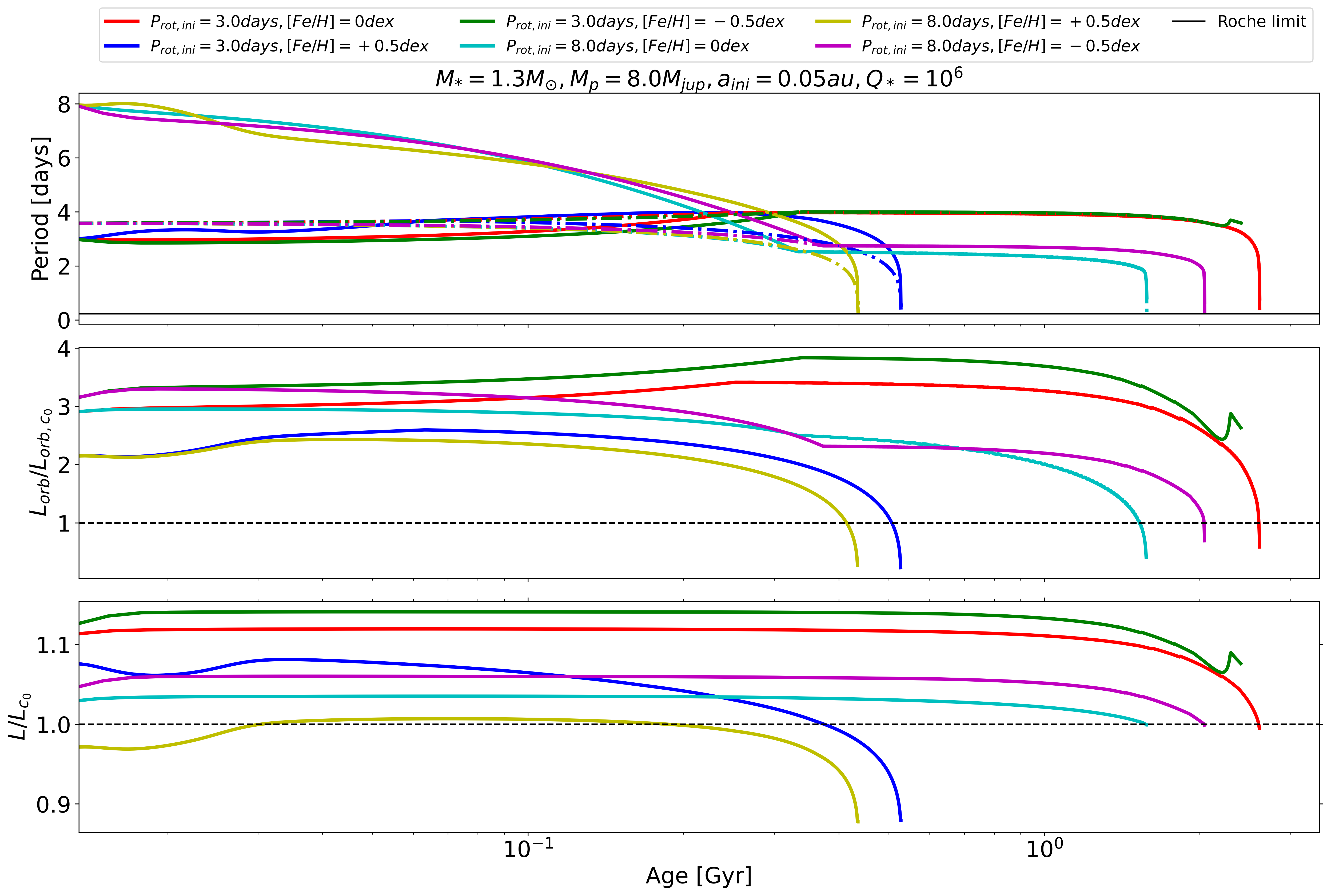}
            \caption{}
            \label{fig2:a}
    \end{subfigure}
    \begin{subfigure}[t]{0.99\textwidth}
            \centering
            \includegraphics[width=0.9\textwidth]{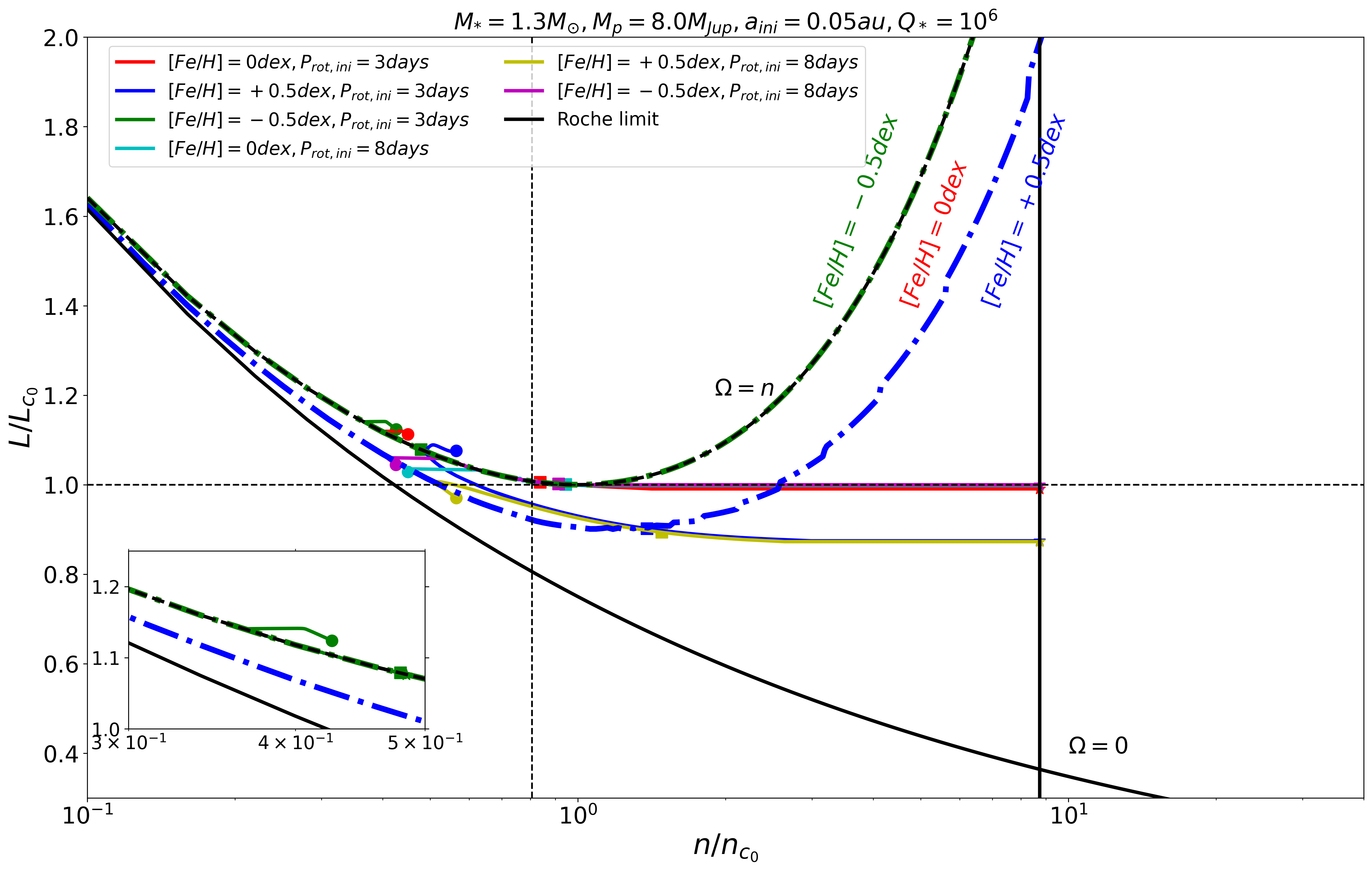}
            \caption{}
            \label{fig2:b}
    \end{subfigure}
\caption{Here, only the stellar mass is changed to 1.3$\,$$M_{\odot}$. Designations are the same as those in Figure \ref{fig:1.1M}.
\label{fig:1.3M}}
\end{figure*}

\subsection{The influence of planetary parameter and tidal quality factor on triggering the double synchronization} \label{subsec:planet mass}

\begin{figure*}
    \centering
    \begin{subfigure}[t]{0.99\textwidth}
           \centering
           \includegraphics[width=0.9\textwidth]{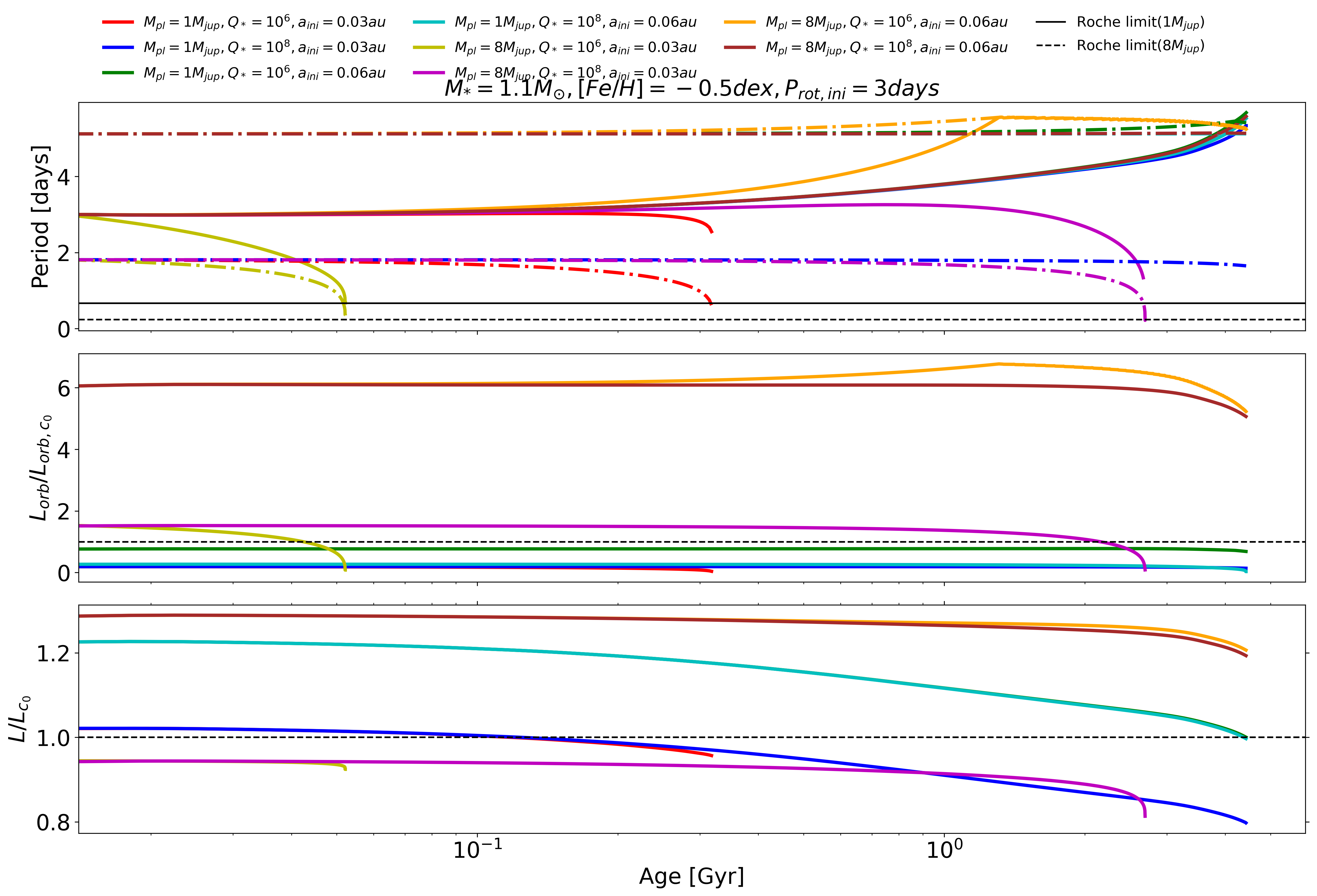}
            \caption{}
            \label{fig3:a}
    \end{subfigure}
    \begin{subfigure}[t]{0.99\textwidth}
            \centering
            \includegraphics[width=0.9\textwidth]{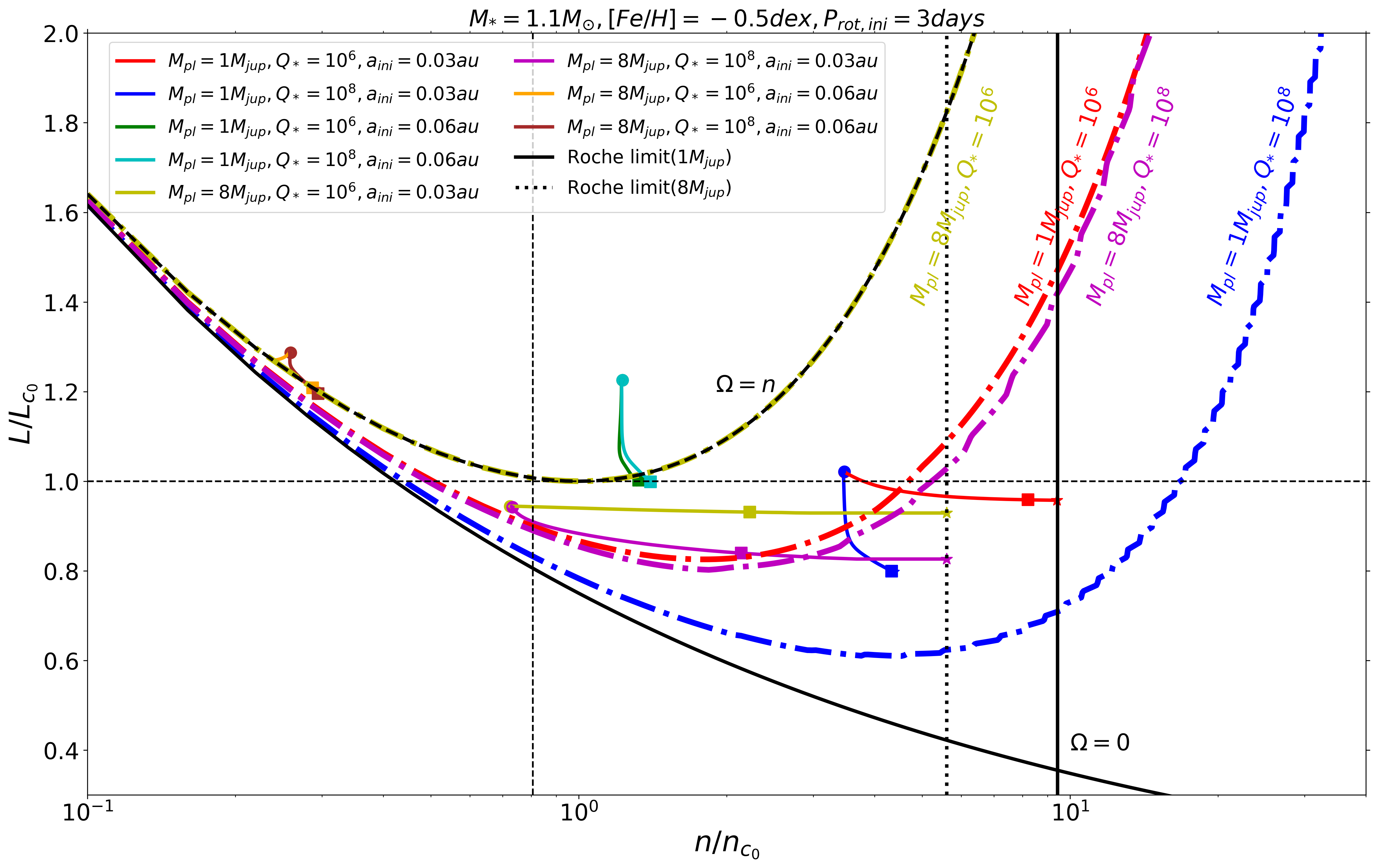}
            \caption{}
            \label{fig3:b}
    \end{subfigure}
\caption{Designations are the same as those in Figure \ref{fig:1.1M}. In panel (a), The star has a mass of 1.1$\,$$M_{\odot}$ and the metallicitiy [Fe/H] = -0.5$\,$dex and the initial stellar rotation period is 3.0$\,$days.
In panel (b), the vertical solid black line and vertical dotted black line represent the position of the Roche limit of 1.0$\,$$M_{\text{J}}$ and 8.0$\,$$M_{\text{J}}$, respectively.
\label{fig:Q}}
\end{figure*}

In this section, we discuss the impact of planet mass, tidal quality factor, and initial orbital semimajor axis on the triggering of the double synchronization mechanism in Figure \ref{fig:Q}. Firstly, from the top panel of Figure \ref{fig:Q}(a), it can be seen that only systems with $a_{\text{ini}}$ = 0.06$\,$au, $Q'_{_*} = 10^6$, and $M_{\mathrm{pl}} = 8.0$\,$M_{\mathrm{J}}$ (orange curve) can undergo the double synchronization mechanism and maintain it. Additionally, it can be observed from the figure that systems with an initial semimajor axis of a = 0.06$\,$au can survive the main sequence stage. The models with an initial semi-major axis of a = 0.03$\,$au survive during the main sequence stage only under the conditions of $Q'{_*} = 10^8$ and $M_{\text{pl}}$ = 1.0$\,$$M_{\mathrm{J}}$. The middle panel of Figure \ref{fig:Q} reveals that systems with a planetary mass of 1.0$\,$$M_{\mathrm{J}}$ have orbital angular momentum below the critical value, while systems with 8.0$\,$$M_{\mathrm{J}}$ have orbital angular momentum above the critical value. As time progresses, for systems with an initial semi-major axis of $a_{\text{ini}}$ = 0.03$\,$au and 8.0$\,$$M_{\mathrm{J}}$, the orbital angular momentum will become lower than the critical value, while for systems with $a_{\text{ini}}$ = 0.06$\,$au and 8.0$\,$$M_{\mathrm{J}}$, the orbital angular momentum will still be above the critical value at the end of the main sequence. The bottom panel indicates that at the beginning of the evolution, only systems with $a_{\text{ini}}$ = 0.03$\,$au and 8.0$\,$$M_{\mathrm{J}}$ have total angular momentum below the critical value. However, as the evolution progresses until the end of the main sequence, only systems with $a_{\text{ini}}$ = 0.03$\,$au will have total angular momentum below the critical value.

In the Darwin diagram of Figure \ref{fig:Q}(b), the equilibrium point trend can generally be divided into two categories: $\romannumeral1$. As $Q'_{_*}$ increases and $M_{\mathrm{pl}}$ decreases, the equilibrium point curve corresponding to $\dot{\Omega}$ = 0 is closer to $\Omega = 0$, indicating that the wind torque plays a more dominant role compared to the tidal torque. $\romannumeral2$. As $Q'_{_*}$ decreases and $M_{\mathrm{pl}}$ increases, the equilibrium point curve corresponding to $\dot{\Omega}$ = 0 is closer to $\Omega = n$, indicating that the tidal torque plays a more dominant role compared to the wind torque.

For the green, light blue, and blue curves in the figure \ref{fig:Q}, the wind torque dominates over the tidal torque, and their stellar rotation speed is faster than the orbital adjustment of the planet, resulting in nearly vertical evolution. The red curve in the figure \ref{fig:Q} quickly drops below the red dash-dotted line as it evolves, indicating that the tidal torque dominates over the wind torque, and the planet is quickly engulfed by the star, resulting in nearly horizontal evolution. The purple and yellow curves quickly evolve past $n_{\mathrm{c_0}}$, where the stronger tidal torque causes them to evolve in an almost horizontal manner. For the orange and brown curves in the figure \ref{fig:Q}, they are located at positions where $\beta$ $\geq$ 1. The planet migrates outward for a period of time, and then the orange curve evolves along the trajectory of $\Omega = \Omega_{\mathrm{sta}}$ until the end of the main sequence stage when it survives. The brown curve evolves towards the equilibrium point, but due to the planet's distance from the star and weaker tidal forces, the trajectory on the Darwin diagram is short and always above the equilibrium point dash-dotted line.

\subsection{The correlation between double synchronization mechanism and initial parameters} \label{subsec:Synchronicity}

\begin{figure*}
\includegraphics[width=0.95\textwidth]{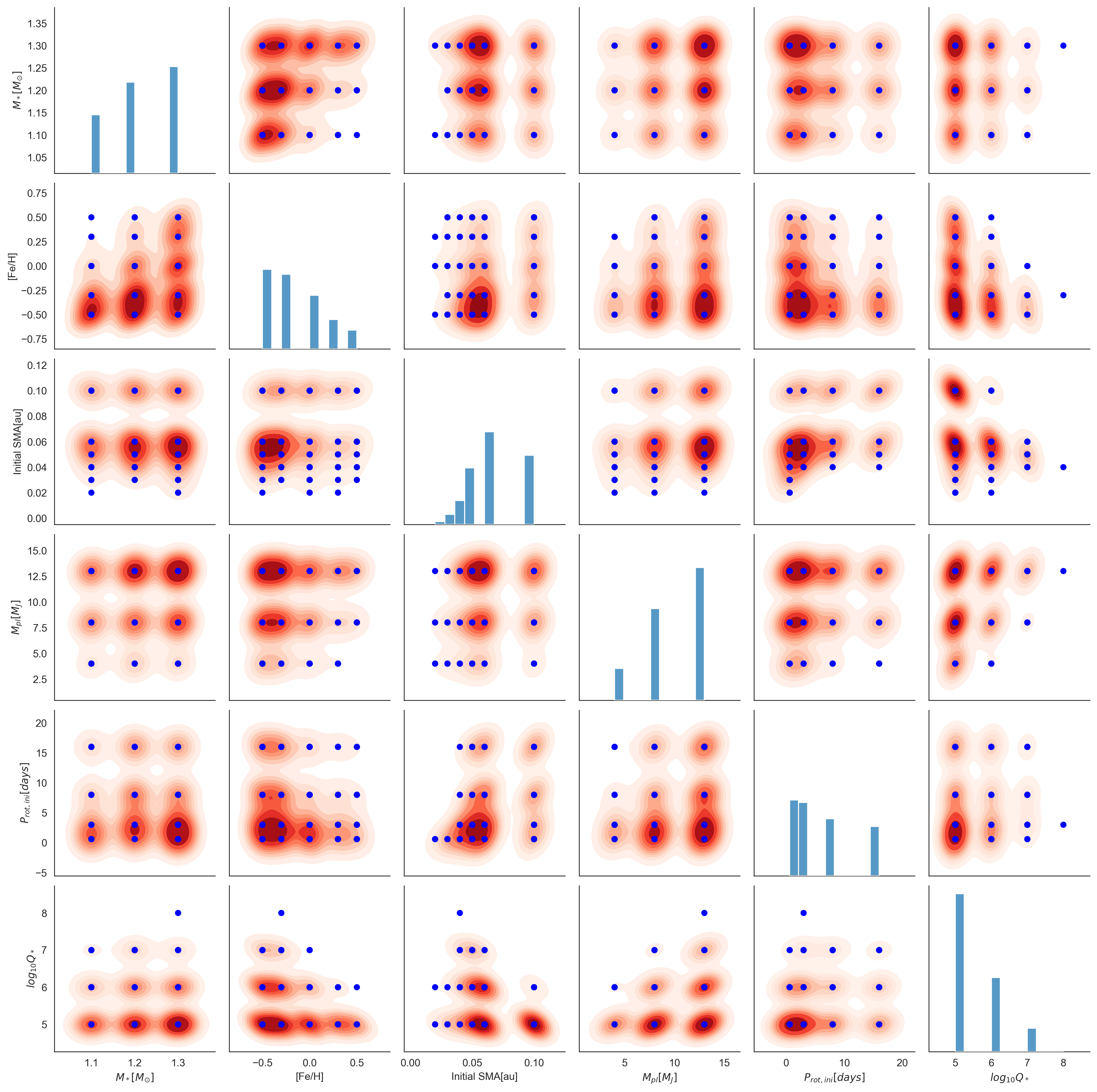}
\caption{blue dots represent samples that can maintain long-term double synchronization under these conditions, while blank areas represent samples that cannot maintain long-term double synchronization. The darker the color of the shaded area, the greater the number of samples at that location. The histogram in the figure represents the distribution of number densities for each initial parameter. There are a total of 425 samples in this part, which is less than 2$\%$ of our total sample size of nearly 25,000. It is important to note that the models we computed form a uniform grid as shown in Table \ref{tab:tab1}. Each shaded region below is marked with blue dots, and the darkness of the shading indicates a larger number of samples in which the double synchronization state occurs. The deeper the color of the shaded region, the larger the range, representing more samples in which the double synchronization state is achieved. The addition of shaded regions is merely for a more intuitive display of the number of samples at different positions in the grid. The area between two grid points, where there is no shading, does not contain any samples. Regions with only blue dots and no shaded portions indicate very few samples at that position, possibly only one or a few.
\label{fig:pyramid}}
\end{figure*}

We present in Figure \ref{fig:pyramid} the initial parameter conditions for a star-planet system to sustain long-term double synchronization. It is noteworthy that the criterion for "long-term double synchronization" we define here is that the star-planet system has undergone this a significant double synchronization phase before the planets are engulfed (we assume that the star and planet maintain co-rotation for more than 10 Myr, then we consider the system to have undergone long-term double synchronization). Firstly, the density distributions of each parameter can be found from the histogram in the Figure \ref{fig:pyramid}. Through our grid calculations, we find that long-term double synchronization can only be sustained in systems with a star mass of 1.1$M_{\odot}$ or greater, an orbital semi-major axis of 0.02$\,$au or greater, a planet mass of 4.0$\,$$M_{\mathrm{J}}$ or greater, and a tidal quality factor $log_{_{10}}\mathcal{Q'}_{_*}$ $\leq$ 8 (Please note that, due to the wide intervals in the parameters of planetary mass, we believe that systems with a planetary mass greater than 1.0$\,$$M_{\mathrm{J}}$ are required for long-term double synchronization to occur. Additionally, it can be seen that we only find a double synchronization system if the tidal quality factor is equal to $10^{8}$, suggesting that long-term double synchronization is more common in systems with $Q'_{_*}$ $\leq$ $10^{8}$). Furthermore, in addition to a peak occurring around an initial semi-major axis of 0.06$\,$au, we can also see from the pairwise relationship diagrams of the initial parameters that larger stellar and planetary masses, as well as smaller values of $log_{_{10}}\mathcal{Q'}_{_*}$, [Fe/H], and the initial stellar rotation period, make it easier for double synchronous orbits to be maintained in the long term. This is because the low-metallicity stars with high mass have thinner convective envelopes that produce weaker angular momentum loss due to magnetic braking. This is reflected in the Darwin diagrams in Figures \ref{fig:1.1M}(b) and \ref{fig:1.3M}(b), where the equilibrium point corresponding to $\dot{\Omega}$ = 0 is closer to the synchronous orbit $\Omega$ = n for $\Omega$ = $\Omega_{\mathrm{sta}}$. On one hand, a weaker magnetic braking results in less stellar angular momentum loss, making it easier for the total angular momentum to remain above the critical value. The state of $\Omega$ $\approx$ n $\approx$ $\Omega_{\mathrm{sta}}$ is allowed for a long time because tidal friction between the star and planet can be ignored and the degree of orbit decay only depends on the weak magnetic braking. Moreover, larger planetary masses and smaller values of $log_{_{10}}\mathcal{Q'}_{_*}$ lead to faster dissipation of tidal energy, which is similar to the previous case and is manifested by an equilibrium point at $\dot{\Omega}$ = 0 that is closer to the synchronous orbit $\Omega = n$ in Figure \ref{fig:Q}(b). 

When the initial stellar rotation period is less than the planetary orbital period, the planet's outward migration will inevitably evolve to the position of $\Omega = n$, implying a higher probability of long-term double synchronization. Additionally, under the same conditions, a faster initial stellar rotation period corresponds to a larger angular momentum, making it easier to achieve $L > L_{\mathrm{c_0}}$. Regarding the initial semi-major axis, if it is too distant, the tidal torque is too weak. To achieve a balance between wind torque and tidal torque, a smaller stellar wind torque is required, resulting in fewer satisfying samples. If the semi-major axis is too close, tidal energy dissipation occurs too rapidly, triggering $n>n_{\mathrm{c_0}}$. Moreover, a high stellar rotation rate is needed to synchronize with the orbital rate, which is evidently challenging. It is noteworthy that an intermediate orbital semi-major axis is more favorable for the long-term maintenance of tidal synchronization. As shown in Figures \ref{fig:1.1M}(b), \ref{fig:1.3M}(b), and \ref{fig:Q}(b), the system remains stable when the evolution is maintained within $\frac{n}{n_{\mathrm{c_0}}} < \frac{3}{4}^{\frac{3}{4}}$.

\section{DISCUSSION} \label{sec:disscussion}

It is highly significant to find the double synchronization system. Therefore, we chose 243 hot Jupiter systems and brown dwarf systems. Based on the results above, we primarily focus on close-in gas giant planetary systems with main sequence stars being F and G$-$type stars \footnote{https://exoplanetarchive.ipac.caltech.edu/index.html}. The stellar mass ranges from 0.8$\,$$M_{*}$ to 1.5$\,$$M_{*}$, and we consider systems with planetary masses above 0.1$\,$$M_\mathrm{{J}}$. The metallicity is within the range of $+0.5 \geq$ [Fe/H] $\geq -0.5$, and the surface gravity is required to be $\text{logg} \geq 3.90$ to ensure that the sample is in the main sequence phase. We select systems with orbital separation within 0.1$\,$au. The parameters of these systems are provided in Table \ref{tab:tab2}. For the majority of planetary systems in our sample, direct measurements of the stellar rotation period are not available. Therefore, we employ the Equation (5) from \citet{2014MNRAS.442.1844B} to calculate the stellar rotation period, where the specific expression is given by $P_{\mathrm{rot}} = \frac{2\pi R_*}{v\sin I_{\mathrm{s}}}\sin i_{\mathrm{orb}}$, the $v\sin I_{\mathrm{s}}$ represents the projection of stellar rotation, and $i_{\mathrm{orb}}$ represents the inclination of the orbit. $R_*$ is stellar radius. Assuming the system is aligned along the line of sight, the inclination of the stellar rotation axis with respect to the line of sight is denoted as $I_{\mathrm{s}} = i_{\mathrm{orb}}$. In order to calculate the convective turnover time($\tau_{\text{cz}}$), we referred to the work of \citet{2021ApJ...912..127S} and utilized the database from \citet{2019A&A...631A..77A} along with the maximum likelihood tool from \citet{2014A&A...561A.125V} to obtain $\tau_{\text{cz}}$. When the stellar masses are higher than 1.5$\,$$M_{*}$, their envelopes are radiative zones, and the influence of metallicity becomes negligible. We calculated the Rossby number $R_{\mathrm{o}} = P_{\mathrm{rot}}/\tau_{cz}$, with the aim is to determine whether the star is in a saturated or unsaturated state. The distribution of these hot Jupiter systems is plotted in the Darwin diagram. We included those samples with a stable solution regime ($\frac{n}{n_{\mathrm{c_{0}}}} < \frac{3}{4}^{\frac{3}{4}}$ and $L > L_{\mathrm{c_0}}$) and fast-rotating stars with $\Omega > n$ from \citet{2015A&A...574A..39D}. These samples are crucial for discussing whether a system is currently in or will experience a double synchronization state in the future.

\subsection{The distribution of planetary systems in the Darwin diagram} \label{subsec:4.1}

\begin{figure*}
\includegraphics[width=0.95\textwidth]{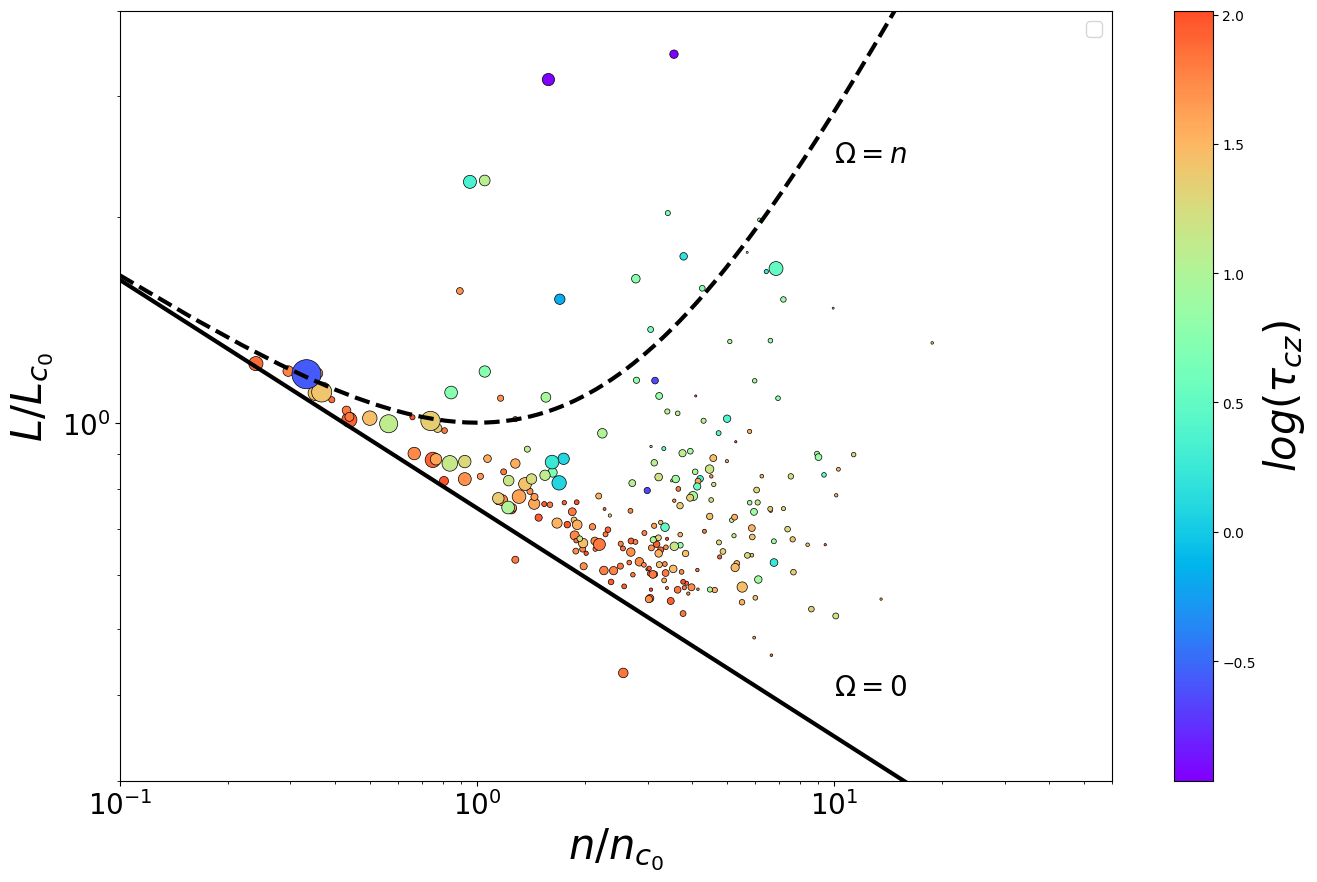}
\caption{In the figure, the circle markers represent observed systems, where the color of the circle markers represents the convective turnover time of the stars, and the size of the circle markers represents the planetary mass. The black dashed line and solid line represent the trajectories for $\Omega = n$ and $\Omega = 0$, respectively.
\label{fig:OB_Darwin}}
\end{figure*}

In Figure \ref{fig:OB_Darwin}, the locations of all the systems in the Darwin diagram, revealing two notable features. This result is consistent with the theoretical model predictions in Figure \ref{fig:pyramid}. In addition, the systems with large convective turnover time are generally closer to the position of $\Omega = 0$ in the diagram. This can be explained by the fact that G-type stars have stronger magnetic braking and slower rotation rates than F-type stars, resulting in lower total angular momentums. Under the same conditions, a larger planetary mass implies a larger orbital angular momentum. The figure also shows that all the planetary mass, stellar mass, and total angular momentum decrease sharply when the orbital velocity is high. This can be attributed to two aspects: on the one hand, when the planet is closer to the star, its orbital angular momentum is smaller than that of a more distant planet under the same conditions. On the other hand, the tidal forces in systems composed of large-mass planets and stars are stronger, leading to faster orbital decay and rapid engulfment of the planet. This makes it difficult to find such systems in close-in orbits.

Finally, for systems with $n/n_{\mathrm{c_0}} > 2$, the planetary mass tends to be smaller. This can be attributed to two reasons: Firstly, \citet{2015A&A...574A..39D} pointed out that for systems with $n/n_{\mathrm{c_0}} > 1.5$, planets with a mass of about 3.0$\,$$M_\mathrm{{J}}$ cannot maintain a stationary state, regardless of the magnetic braking efficiency of their host stars. This suggests that massive planets in close proximity find it difficult to remain stable, as tidal forces can lead to their rapid engulfment. Secondly, a larger planetary mass is usually associated with a larger critical frequency ($n_{\mathrm{c_0}}$), which to some extent favors smaller values of $n/n_{\mathrm{c_0}}$ for systems with larger planetary masses.

\subsection{The Phases of long-term double synchronization Experienced by Planetary Systems} \label{subsec:4.2}
Through the analysis of the equilibrium states of planetary systems, we can infer whether they can maintain long-term double synchronization. As these systems evolve, the planets in a state of long-term double synchronization may eventually stay at this state, while systems currently non-synchronized could potentially enter into such a state with the evolution.

\begin{figure*}
    \centering
    \begin{subfigure}[t]{0.48\textwidth}
        \centering
        \includegraphics[width=\textwidth]{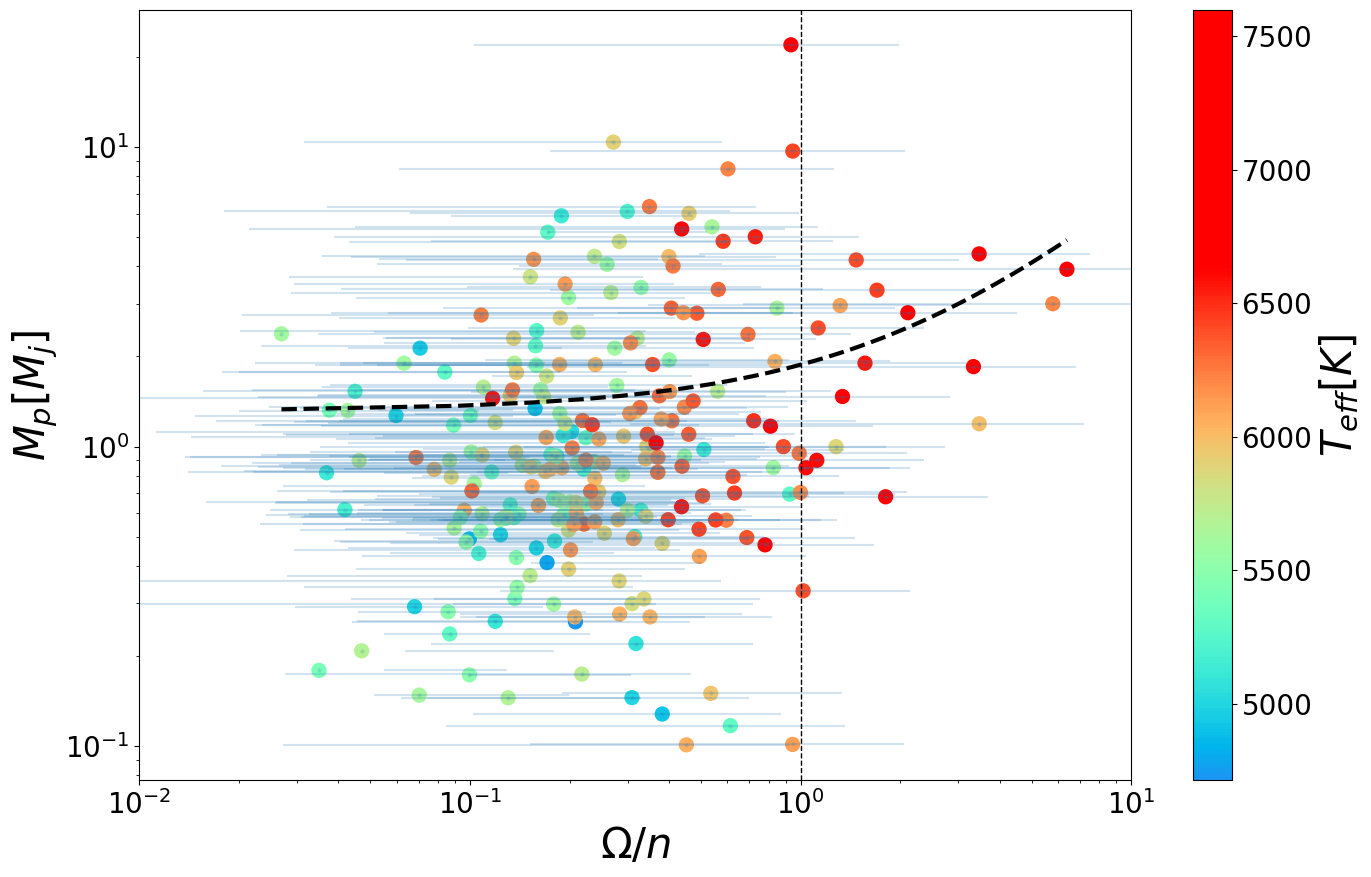}
        \caption{}
        \label{fig6:a}
    \end{subfigure}
    \hfill 
    \begin{subfigure}[t]{0.48\textwidth}
        \centering
        \includegraphics[width=\textwidth]{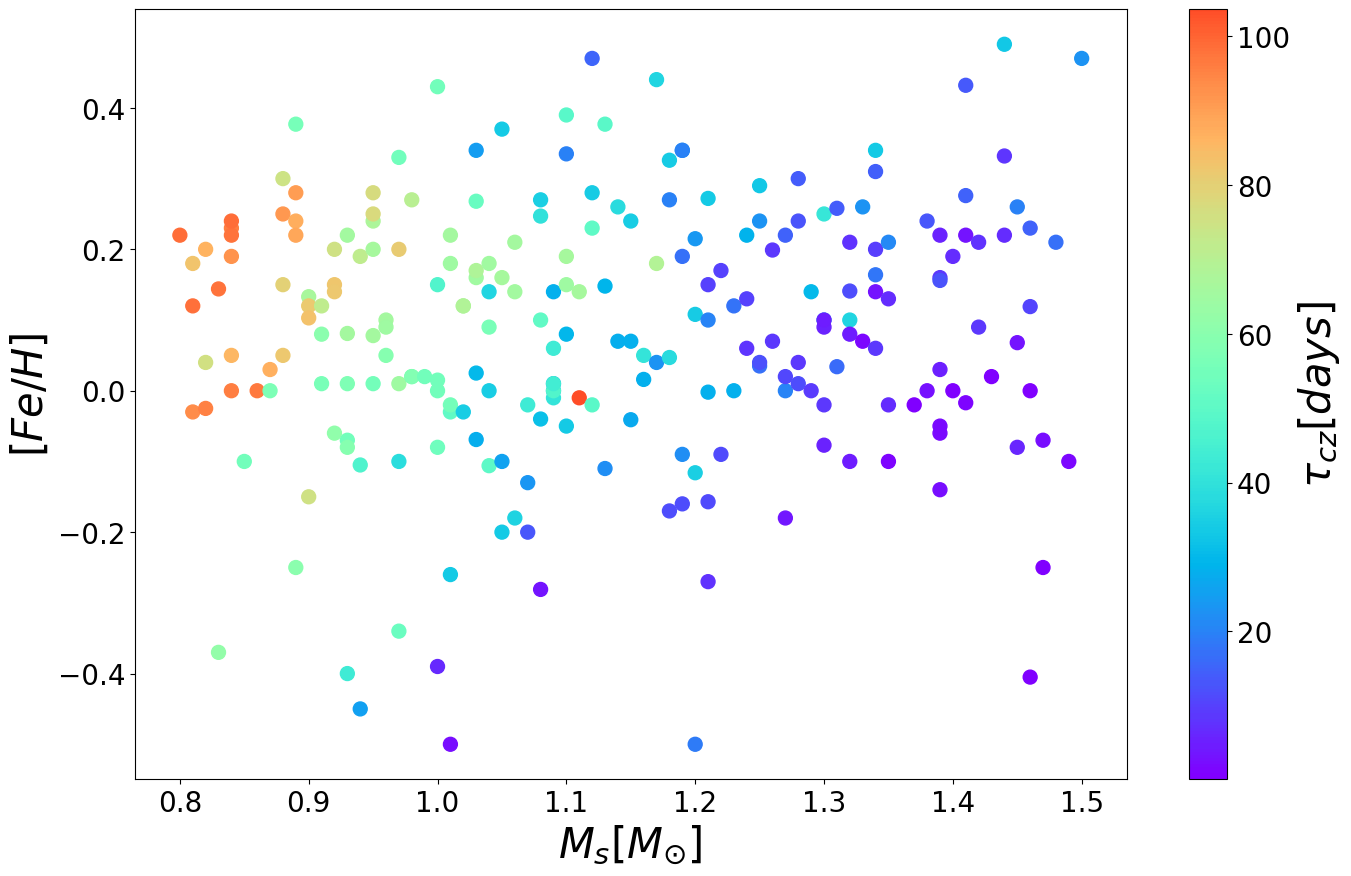}
        \caption{}
        \label{fig6:b}
    \end{subfigure}
    \caption{In panel (a), the plot illustrates the relationship between $\Omega/n$, planetary mass, and effective temperature. Colored dots represent observed systems, while the black dashed line represents a linear fit to $\Omega/n$ and planetary mass. The light blue thin lines added to each sample represent error bars. In panel (b), figure depicts the relationship among stellar mass, metallicity, and the convective turnover time, colored dots represent observed systems.}
    \label{fig:OB}
\end{figure*}

We show the relationship between $\Omega/n$, planetary mass and effective temperature in Figure \ref{fig:OB}(a) for our selected observational samples. It can be seen from the figure that $\Omega/n$ and $M_{\text{pl}}$ have a positive correlation, and the smaller the planetary mass and $\Omega/n$, the lower the effective temperatures of the stars. It can also be found that in the region where $\Omega/n \geq 0.3$, the effective temperatures are above 6000$\,$K. This is because the stars in the low-temperature regions have stronger magnetic braking because of the thicker convective envelopes, and their rotation rates are often slower than those in the high-temperature regions. This leads to two consequences. First, there is a smaller $\Omega/n$ for stars in the low-temperature. Second, the planets with large mass can be quickly engulfed by their host star with lower effective temperatures due to a rapid transfer of the planetary angular momentum to star. This also makes it very difficult for us to find outward-migrating ($\Omega/n \geq 1$) samples in the low-temperature region.

We also show the relationship between stellar mass, metallicity and convective turnover time in Figure \ref{fig:OB}(b). We can clearly see that the convective turnover time decreases with the increase of stellar mass, and there is weak correlation between metallicity and convective turnover time. At a given stellar mass, the convective turnover times increase with increasing metallicity.

\begin{figure*}
	\centering
	\begin{subfigure}[t]{0.48\textwidth}
		\centering
		\includegraphics[width=\textwidth]{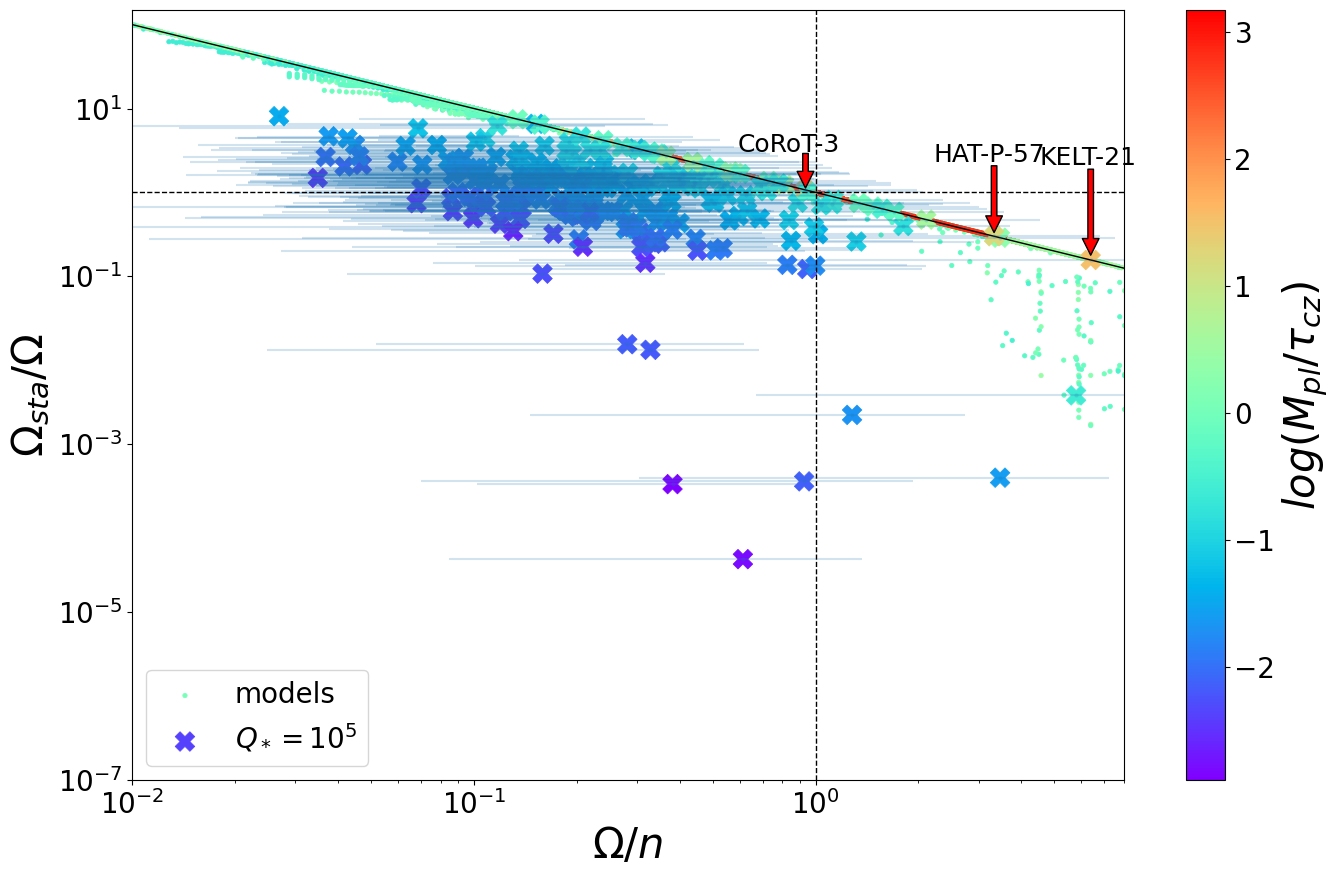}
		\caption{}
		\label{fig7:a}
	\end{subfigure}
	\hfill 
	\begin{subfigure}[t]{0.48\textwidth}
		\centering
		\includegraphics[width=\textwidth]{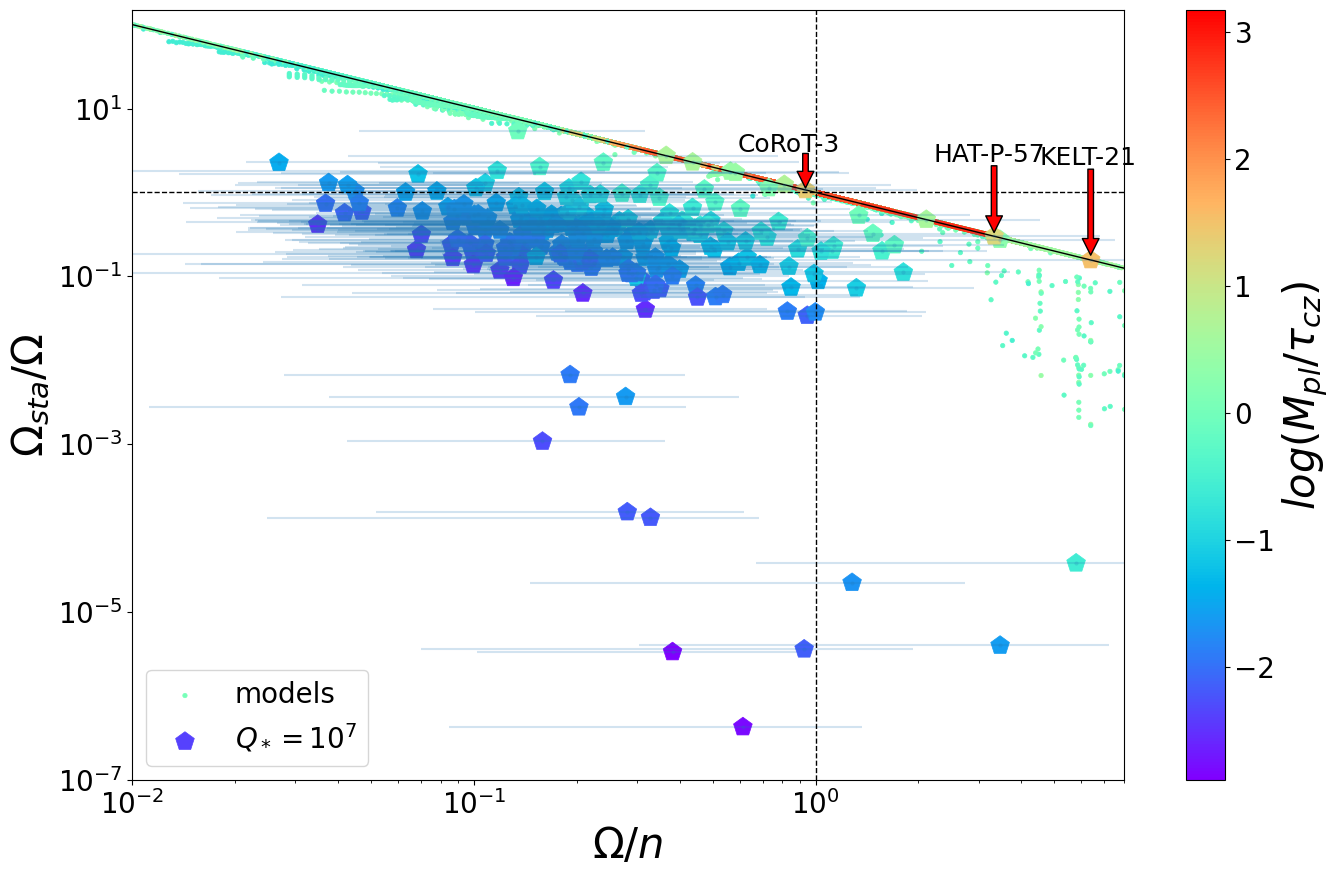}
		\caption{}
		\label{fig7:b}
	\end{subfigure}
	\begin{subfigure}[t]{0.48\textwidth}
		\centering
		\includegraphics[width=\textwidth]{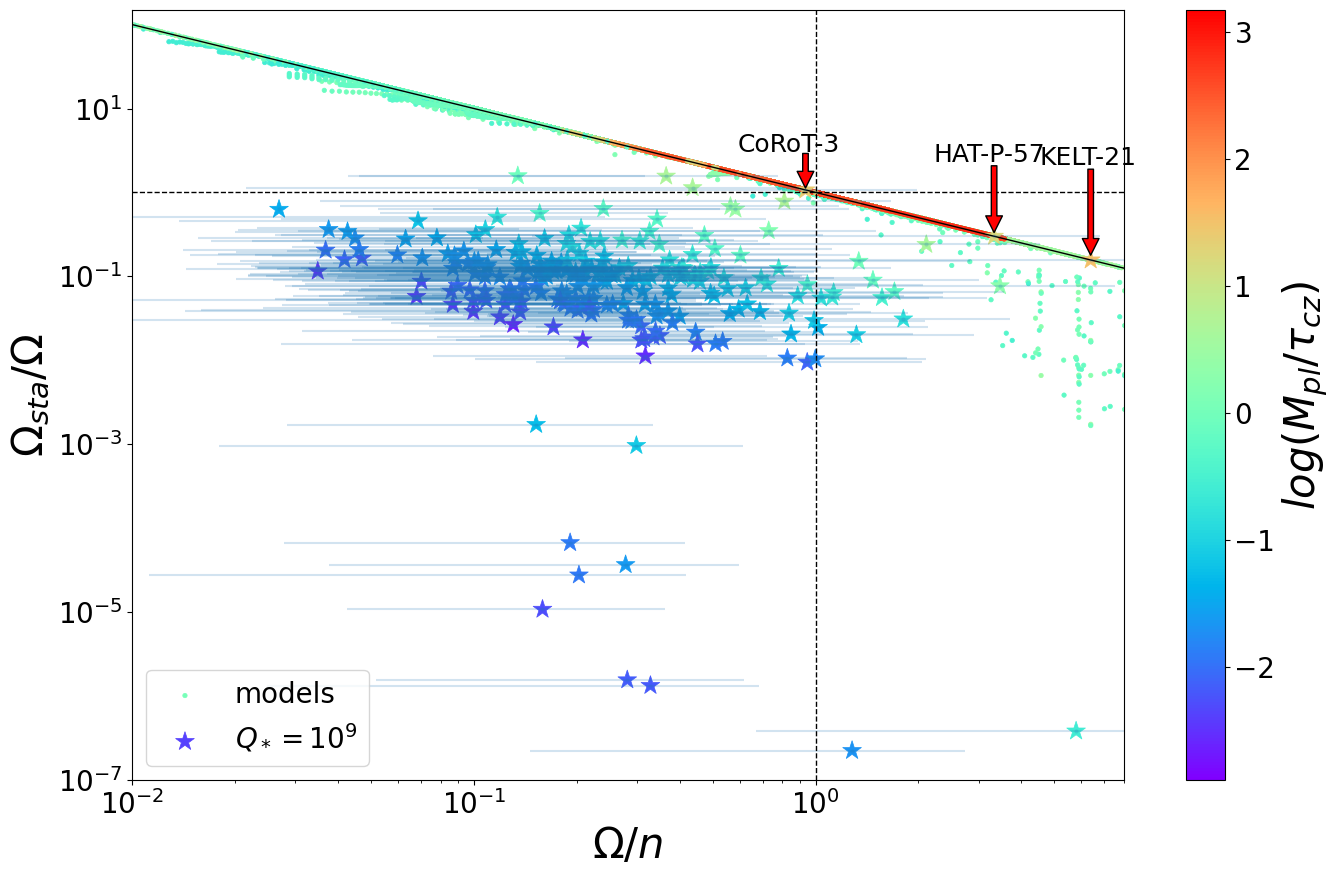}
		\caption{}
		\label{fig7:c}
	\end{subfigure}
\caption{The plot illustrates the distribution of $\Omega/n$ and $\Omega_{\mathrm{sta}}$/$\Omega$, where crosses(panel (a)), diamonds(panel (b)) and pentagram(panel (c)) symbols represent the $\Omega_{\mathrm{sta}}$ values calculated for 243 observational systems with tidal quality factors $Q'_{_*}$ of $10^5$, $10^7$, and $10^9$, respectively. The black horizontal and vertical dashed lines represent $\Omega_{\mathrm{sta}}$/$\Omega$ = 1 and $\Omega/n$ = 1. The colored circles represent data points from the entire main-sequence evolution of 425 theoretical models that have undergone long-term double synchronization. The black solid line represents the curve where $\Omega_{\mathrm{sta}}=n$. The systems marked with red arrows in the figure are those with tidal quality factor $Q'_{_*}$ ranging from $10^5$ to $10^9$, all situated at $\Omega{\mathrm{sta}} \approx n$. The light blue thin lines added to each sample represent error bars.
\label{fig:Lsta}}
\end{figure*}

In Figure \ref{fig:Lsta}, we present the relationship between $\Omega/n$ and $\Omega_{\mathrm{sta}}$/$\Omega$ for the 425 theoretical models that have experienced long-term double synchronization and 243 observational data with tidal quality factors $Q'_{_*}$ of $10^5$, $10^7$, and $10^9$. We find that on a logarithmic scale, those systems that have experienced a long-term double synchronization state are situated near the curve of $\Omega_{\mathrm{sta}}=n$. This means that for systems that have undergone long-term double synchronization states, the equilibrium state's $\Omega_{\mathrm{sta}}$ is very close to the orbital rate n for most of the time in life. The figure also shows that some discrete points with smaller $M_{\mathrm{pl}}/\tau_{\text{cz}}$ values are below the curve of $\Omega_{\mathrm{sta}}=n$. Considering the distribution range of $\Omega_{\mathrm{sta}}$/$\Omega$, we assume points with $(\Omega_{\mathrm{sta}}-n)/\Omega \geq 0.01$ as discrete points, which is less than 0.5\% of the total sample. Moreover, the discrete points with $\Omega/n < 1$ occur when the planet is almost to be engulfed (in the last 1\% of the entire evolutionary stage), where tidal torques dominate. The discrete points with $\Omega/n > 1$ are at the early phase of the evolution when the planet starts to migrate outward (in the first 1\% of the entire evolutionary stage). For points with $(\Omega_{\mathrm{sta}}-n)/\Omega \leq 0.01$, we consider as the situation of $\Omega_{\mathrm{sta}} \approx n$, and this sample can be understood as falling on the curve $\Omega_{\mathrm{sta}}=n$. The figure illustrates that long-term double synchronization states require larger $M_{\mathrm{pl}}/\tau_{\text{cz}}$, which corresponds to shorter convective turnover times and larger planetary masses, consistent with the findings presented in Section \ref{sec:result}. From the figure, we can also learn that even though these systems that have undergone long-term double synchronization states cannot guarantee to maintain this state throughout the entire main sequence of the star, most of them can still maintain $\Omega_{\mathrm{sta}} \approx n$ in the evolution process. This provides a new approach to discover long-term double synchronization systems. Using the condition of $\Omega_{\mathrm{sta}}=n$, we cannot only find systems that are currently in this state, but also  the systems that have experienced this state in the past ($\Omega/n < 1$), and systems that may experience this state in the future ($\Omega/n > 1$). For example, it is evident that there are three systems with larger $M_{\mathrm{pl}}/\tau_{\text{cz}}$ values that fall on the curve $\Omega_{\mathrm{sta}}=n$ regardless of the $Q'_{_*}$ value, though the planetary system’s positions are closer to the curve of $\Omega_{\mathrm{sta}}=n$ if $Q'_{_*}$ is smaller. These three systems are CoRoT-3, HAT-P-57, and KELT-21. Among them, CoRoT-3 is near $\Omega/n$=1, and this system may currently be in a long-term double synchronization state. HAT-P-57 and KELT-21 are located in the area where $\Omega/n > 1$, and as they evolve, they will maintain a long-term double synchronization state if these two systems can achieve orbital synchronization during the main sequence.

In the regions of $\Omega/n < 1$, many points with $Q'_{_*}$ = $10^5$ and some with $Q'_{_*}$ = $10^7$ also fall on the curve. This indicates that if the initial state allows the system to experience orbital synchronization, then the synchronization state that occurred in the past can also be maintained for a long time. If we can accurately measure these $Q'_{_*}$ values in the future, for planets that fall on the curve $\Omega_{\mathrm{sta}}=n$, we can more accurately determine the stages of long-term double synchronization states experienced by the system. The vast majority of systems that fall below this curve can be excluded, even though many of them are in stable states and near synchronous orbits on the Darwin diagram. However, this synchronous orbit will be immediately broken as they evolve.

\begin{figure*}
\includegraphics[width=0.95\textwidth]{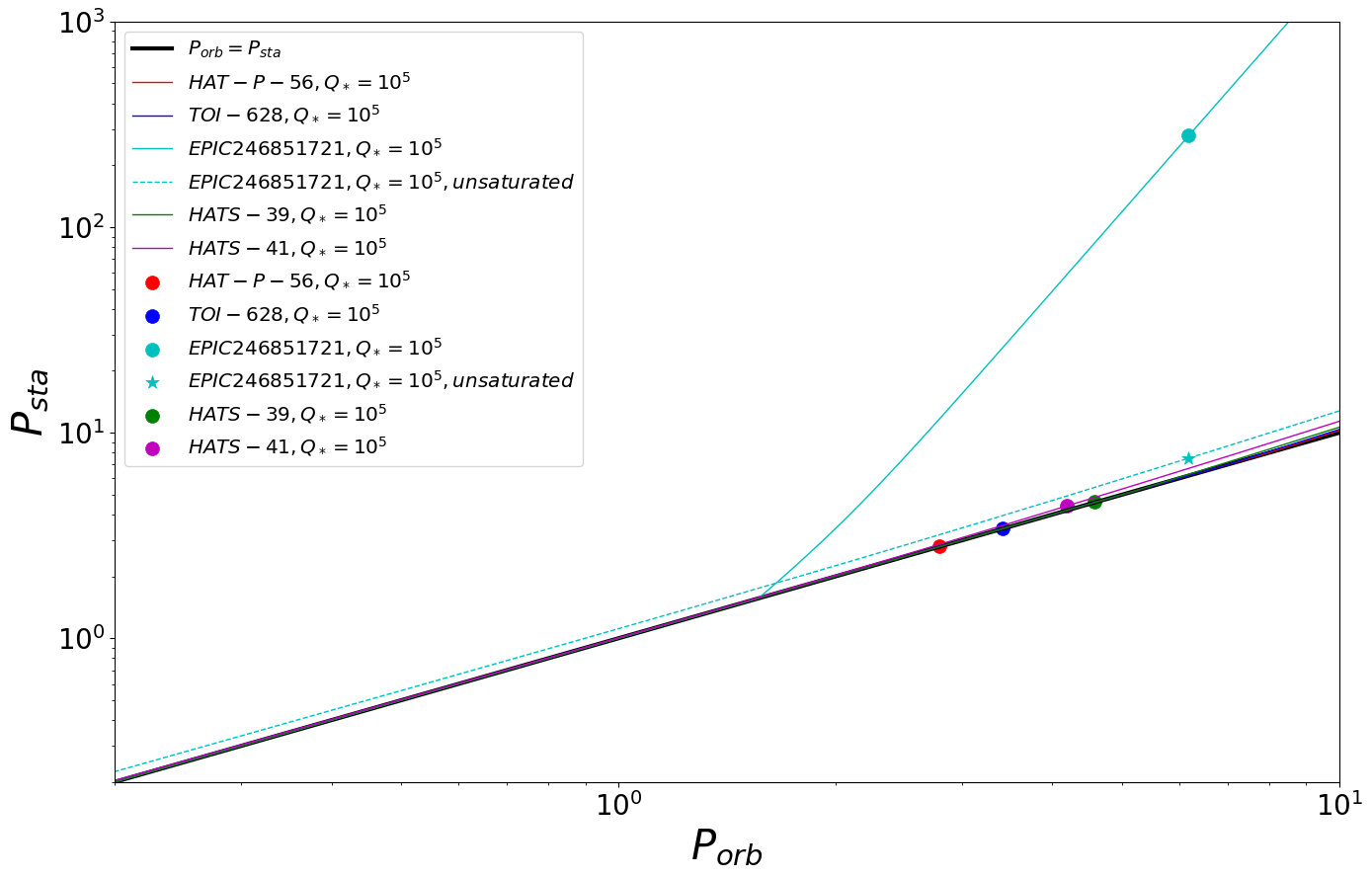}
\caption{The figure shows the relationship between orbital period $P_{\text{orb}}$ and stellar rotation period corresponding to torque balance $P_{\text{sta}}$ for five observational systems: HAT-P-56, TOI-628, EPIC 246851721, HATS-39, and HATS-41. The black bold line indicates the position where $P_{\text{orb}}$=$P_{\text{sta}}$, and all the colored solid lines represent the relationship between $P_{\text{sta}}$ and $P_{\text{orb}}$ for each system. The cyan dashed line is our hypothetical relationship for EPIC 246851721 when it is in an unsaturated state. The circle and pentagram symbols denote the current positions of the systems.
\label{fig:Psta}}
\end{figure*}

Orbital evolution can lead to changes in $\Omega_{\mathrm{sta}}$, which raises the question of whether systems currently at $\Omega_{\mathrm{sta}} \approx n$ will evolve to $\Omega_{\mathrm{sta}} < n$, and if systems at $\Omega_{\mathrm{sta}} < n$ will evolve to reach $\Omega_{\mathrm{sta}} \approx n$. To illustrate this issue, we searched for systems within an orbital period range of 0.2 to 10.0$\,$days in which some maintain $\Omega_{\mathrm{sta}} < n$ while the rest are $\Omega_{\mathrm{sta}} = n$. We have identified five systems exhibiting this particular characteristic, which only manifests when the tidal quality factor $Q'_{_*}$ = $10^5$. Figure \ref{fig:Psta} shows the relationship between $P_{\text{orb}}$ and $P_{\text{sta}}$, where HAT-P-56 and EPIC 246851721 are currently in the state of $\Omega/n > 1$, indicating outward orbital migration. Interestingly, EPIC 246851721 is in a saturated state ($R_{\text{o}} < 0.14$), and as $P_{\text{orb}}$ increases, $P_{\text{sta}}$ deviates further from $\Omega_{\mathrm{sta}} \approx n$. However, as the stellar rotation period increases through evolution, the system will transit to an unsaturated state before orbital synchronization, where systems are at $\Omega_{\mathrm{sta}} < n$, this prevents long-term double synchronization. For HAT-P-56, in cases where orbital period $P_{\text{orb}}$ is greater than 8.0 days, $P_{\text{sta}}$ gradually moves away from $\Omega_{\mathrm{sta}} = n$. If synchronization occurs within $P_{\text{orb}} < 8$$\,$days, this long-term synchronized state will also be experienced. TOI-628, HATS-39, and HATS-41 are currently in a state where $\Omega/n < 1$, and HATS-41's orbit is almost synchronized. However, HATS-41 is found to be at $\Omega_{\mathrm{sta}} < n$, indicating that this synchronization will not be maintained in the long term. The situations for TOI-628 and HATS-39 are more uncertain. If orbital synchronization occurred during a past evolutionary stage at $\Omega_{\mathrm{sta}} \approx n$, the long-term double synchronization would be experienced. Otherwise, even if orbital synchronization has occurred, it will not be sustained in a long time.

\section{CONLUSION} \label{sec:conlusion}

We used MESA for modeling and considered \citet{2015ApJ...799L..23M} magnetic braking model based on \citet{2015A&A...574A..39D}. In this case, we were able to discuss in detail the relationship between magnetic braking and stellar mass and metallicity, and explore the different effects of stellar mass and metallicity on system stability. Due to the complexity of the model, we solved for the stellar rotation rate $\Omega_{\mathrm{sta}}$ at the balance between the wind torque and tidal torque for various parameter values. We extensively discussed the relationship between the stellar rotation period, planetary orbital period, $\frac{L_{\mathrm{orb}}}{L_{\mathrm{orb,c_0}}}$ and $\frac{L}{L_{\mathrm{c_0}}}$ as a function of time for various initial parameters. We found that long-term double synchronization is more easily achieved for higher-mass stars, lower metallicity, smaller $Q'_{_*}$, larger planetary mass, moderate orbital distance, and faster initial stellar rotation rate. In addition, we discussed the evolution of $\Omega_{\mathrm{sta}}$ as a function of $\frac{n}{n_{\mathrm{c_{0}}}}$ for various parameter values. We found that for lower metallicity, higher stellar mass, higher planetary mass, and smaller tidal quality factor $Q'_{_*}$, the equilibrium curve of $\Omega_{\mathrm{sta}}$ is closer to $\Omega$ = n, and vice versa. When the equilibrium curve of $\Omega_{\mathrm{sta}}$ coincides with $\Omega$ = n, the system will maintain long-term double synchronization under the stable solutions proposed by \citet{1981A&A....99..126H} and \citet{2015A&A...574A..39D} ($\frac{n}{n_{\mathrm{c_{0}}}} < \frac{3}{4}^{\frac{3}{4}}$ and $L > L_{\mathrm{c_0}}$).

We calculated uniform grids for nearly 25,000 models and found that 425 systems can undergo and maintain a long-term double synchronous phase. It is believed that in certain conditions, a star-planet system can maintain a double synchronous state for a long time, even if the planet is a poor metal star. Generally, when the planetary mass is larger and the tidal quality factor is smaller, tidal interactions tend to be stronger, and planet engulfment occurs faster. However, when the planetary orbit distance is moderate and the stellar rotation period is equal to the planetary orbit period, weaker magnetic braking can maintain double synchronization in stars with thinner convective envelopes. This mechanism may make these planets easier to survive. We conducted a quantitative analysis of double synchronization and found that:

(\romannumeral1) For stars, systems that maintain double synchronization for a long time tend to occur when the star mass is 1.1$\,$$M_{\odot}$ or higher. And as their mass increases, the upper limit of metallicity is also higher. As for the initial rotation period of the star, if it is less than the orbital period of the planet, the exchange of angular momentum between them will be faster, and therefore a faster stellar rotation period also tends to maintain double synchronization more easily.

(\romannumeral2) For planets, in most cases, when the initial semi-major axis $a$ is around 0.06$\,$au, the planet mass $M_{\mathrm{pl}}$ $\geqslant$ 4.0$\,$$M_{\mathrm{J}}$, and $log_{10}\mathcal{Q'}_{*}$ $\leqslant$ 7, double synchronization will be more easier to occur. The recent studies indicate that the logarithm of the tidal quality factor $log_{10}\mathcal{Q'}_{*}$ for solar-like stars with orbital periods less than 15.0$\,$days falls within the range of 5.7 to 7.5. \citep{2023MNRAS.524.5575P} This means that when the convective layer of the host star is thin and the stellar rotation period is close to the planetary orbit period, the system is easier to be in a long-term double synchronization, making it easier for close-in and massive planets to survive during the main sequence of the star.

We surveyed a sample of 243 hot Jupiter and brown dwarf systems, with the primary stars being F-type and G-type stars that have relatively complete observational parameters. We calculated the convective turnover timescale, $\tau_{\text{cz}}$, for each star and, in conjunction with theoretical models, found that for the majority of the main sequence stage, the equilibrium position $\Omega_{\mathrm{sta}}$ of long-term double synchronization systems is close to the current orbital rate n, regardless of whether the system is in a co-rotating state. We also discovered that in the regions near $\Omega/n \geq 1$, the systems CoRoT-3, HAT-P-57, and KELT-21 are positioned at $\Omega_{\mathrm{sta}} \approx n$ regardless of the value of $Q'_{_*}$. Our work provides theoretical constraints for identifying observational samples in a double synchronization state, and we can find not only systems that may currently be in a long-term double synchronization state but also predict systems that have experienced or may experience this phenomenon in the past and future. As the tidal quality factor $Q'_{_*}$ is measured more precisely in the future, more and more systems that have gone through this phase will be discovered.

software:MESA R11554 \citep{ 2011ApJS..192....3P,2013ApJS..208....4P,2015ApJS..220...15P,2018ApJS..234...34P,2019ApJS..243...10P}

\section*{Acknowledgements}
I am grateful to the reviewers for their insightful feedback and valuable contributions to this manuscript. This work is Supported by the Strategic Priority Research Program of Chinese Academy of Sciences, Grant No. XDB41000000. J. Guo knowledges support from National Natural Science Foundation of China (Nos.11973082 and 42305136). The National Key R$\&$D Program of China, Grant No. 2021YFA1600400/2021YFA1600402. The Natural Science Foundation of Yunnan Province (No. 202201AT070158)

\section{Data availability}
The data underlying this article will be shared on reasonable request to the corresponding author.

\bibliographystyle{mnras}
\bibliography{example.bib} 
\appendix
\section{Comparison between the differential rotation and the solid body rotation models of the Sun.} \label{subsec:sun}
In this section, we first present the infinitesimal element of the stellar moment of inertia:

\begin{equation}
\mathrm{dI} = 2\rho(r)r^4(sin\theta)^3 dr d\theta d\phi
\label{eq27}
\end{equation}

In the equations, r, $\theta$, and $\phi$ are the distances in spherical coordinates from a point to the origin. The angles represent the inclination of the line connecting the point and the origin with respect to the z-axis, and the projection of this line onto the xy-plane with respect to the x-axis. The range of r is [0,$R_{\odot}$], $\theta$ varies within [0, $\pi$/2] and $\phi$ ranges from [0, $2\pi$].

\begin{figure*}
\includegraphics[width=0.95\textwidth]{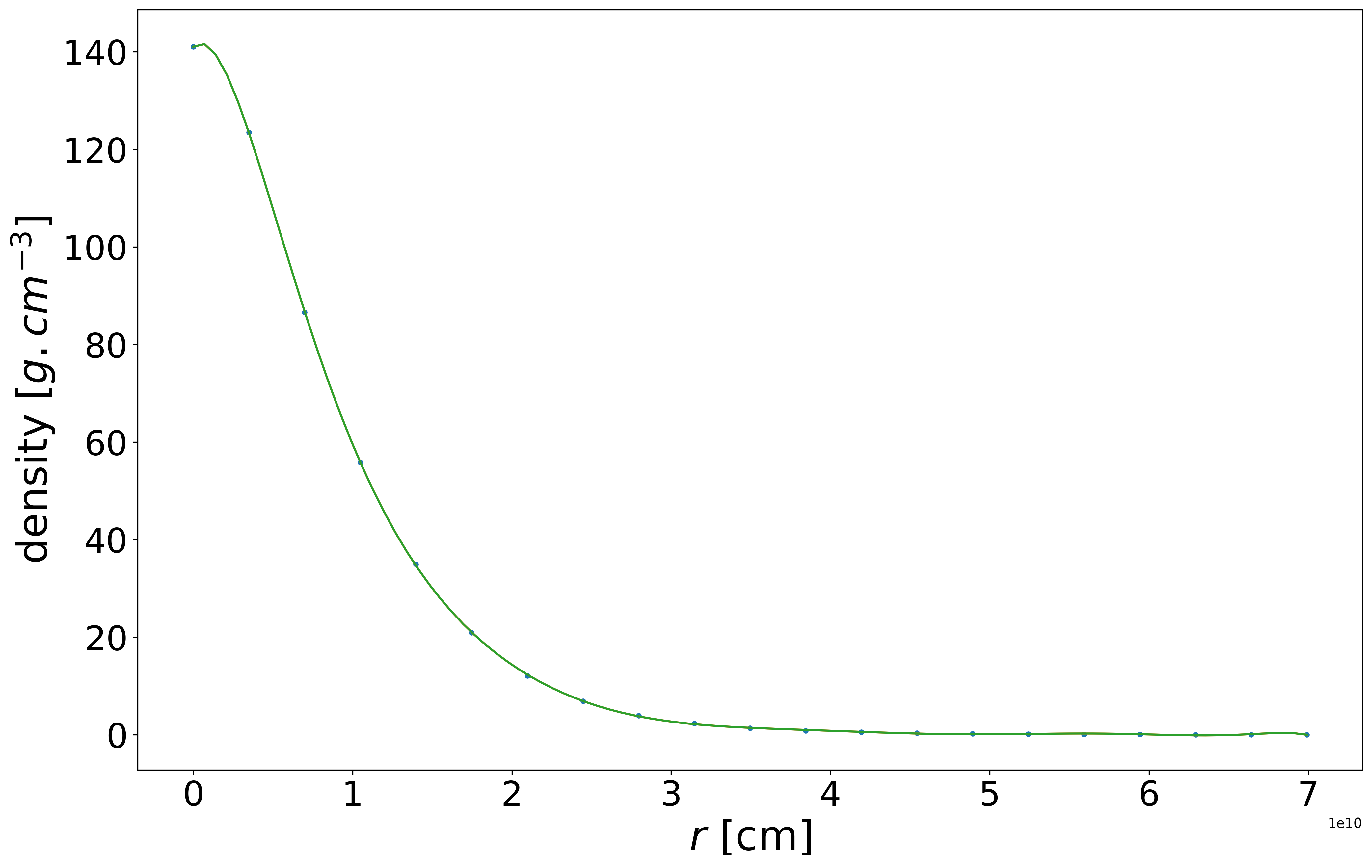}
\caption{The figure depicts the relationship between the position from the center to the interior of the Sun and its density. The blue dots represent asteroseismic data provided by \citet{1994ApJ...432..417D}, while the green solid line represents the polynomial fitting results we obtained.   \label{fig:I}}
\end{figure*}

Continuing, we provide the relationship between solar density and radius derived through asteroseismology by \citet{1994ApJ...432..417D} in Figure \ref{fig:I}. For the description of the Sun's differential rotation, we employed the formula proposed by \citet{1990ApJ...351..309S}.

\begin{equation}
\Omega = A+B(cos\theta)^2+C(cos\theta)^4
\label{eq28}
\end{equation}

$\Omega$ represents the angular velocity in degrees per day, with A, B, and C as constants. The current accepted values for these constants are:
A = 14.713 degrees/day, 
B = -2.396 degrees/day, 
C = -1.787 degrees/day

Finally, we utilized the aforementioned equation to calculate the normalized moments of inertia under two scenarios (sun rotation with differential rotation, using the equatorial rotation rate in the solid body model). The results are presented in the table \ref{table3:number}. We find that the impact of differential rotation results in a difference of normalized moments of inertia of just under 5\%, a negligible effect.

\begin{table}

\caption{The normalized moments of inertia}
\centering
\begin{tabular}{|c|c|c|}
    \hline
    & solid body model & differential rotation model \\

    \hline
    Value & 0.0760 & 0.0727 \\
    \hline
\end{tabular}
\label{table3:number}
\end{table}

\section{Methods for Finding Roots of High-Degree One-Dimensional Equations} \label{subsec:py}

We use the $\mathtt{root\_scalar}$ function in the $\mathtt{SciPy}$ library of $\mathtt{Python}$ to obtain numerical solutions to the equation. $\mathtt{root\_scalar}$ is one of the functions in $\mathtt{SciPy}$ used to find roots of one-dimensional scalar functions. It uses various algorithms to search for the roots of the given function and returns the objects of these roots. In this case, we set $\mathtt{method = 'brentq'}$, maximum iterations $\mathtt{maxiter = 1000}$, and the root interval $\mathtt{bracket = [0, n]}$. $\mathtt{'brentq'}$ refers to the Brent algorithm, it combines the robustness of the bisection method with the fast convergence of Newton's method, making it one of the efficient algorithms for solving the zero-point problem of a univariate function. From equation 4, we can see that $\Omega > 0$ and $(n - \Omega)>0$ must hold for the equation to be equal to zero and satisfy physical conditions. Therefore, the root interval that satisfies physical conditions must be between $\mathtt{[0, n]}$.

\section{Model Comparison} \label{subsec:comp}

\begin{figure*}
\includegraphics[width=0.95\textwidth]{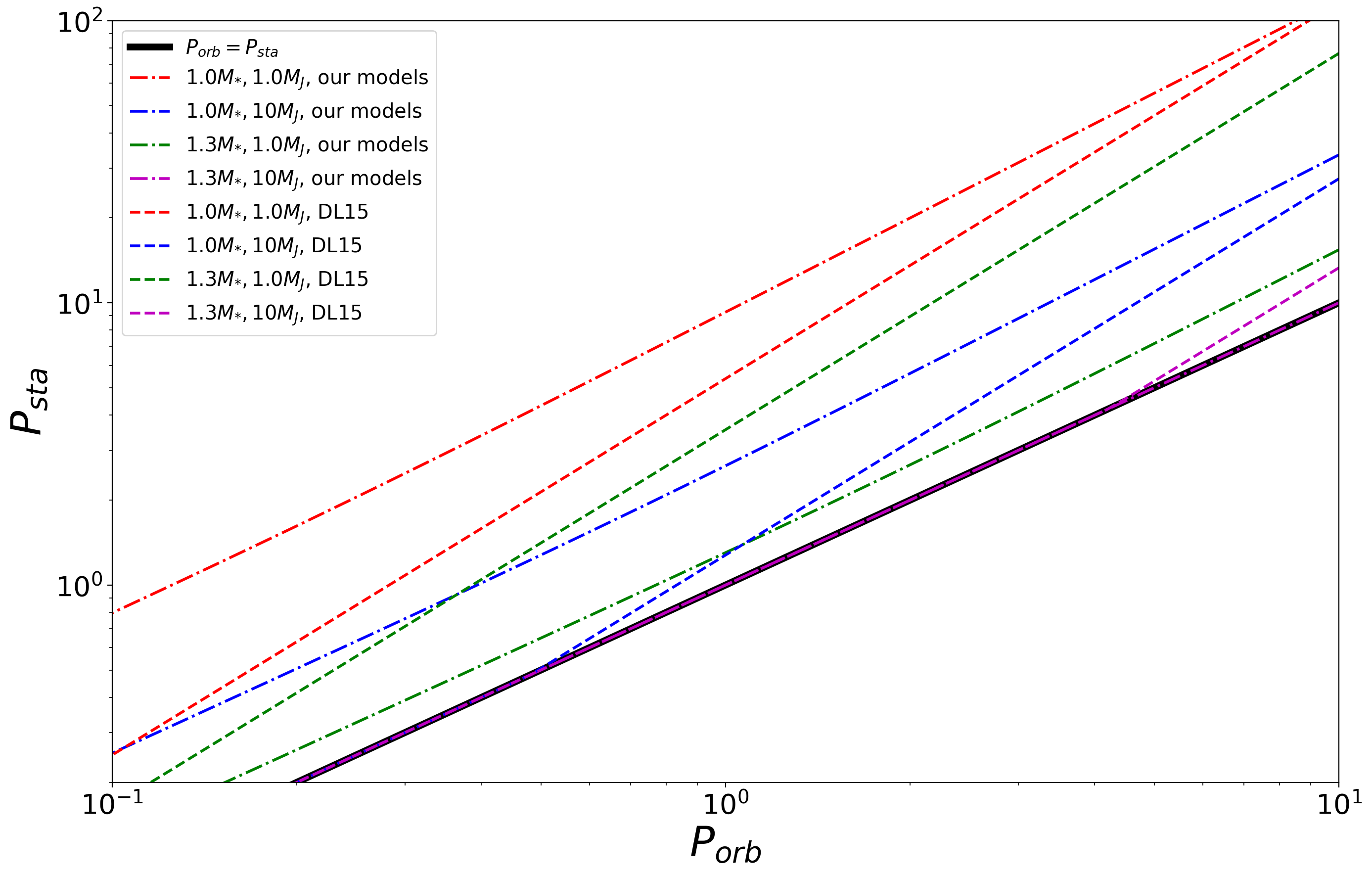}
\caption{Similar to Figure \ref{fig:Psta}, we use dash-dotted lines to represent the results of our model and dashed lines to represent the results of the DL15 model.
\label{fig:C1}}
\end{figure*}

This section aims to compare the results obtained from our improved model with those from the DL15 model (We refer to \citet{2015A&A...574A..39D} model as the DL15 model). Firstly, our model integrates a more intricate magnetic braking model, which takes into account the influence of metallicity. The impact of metallicity is significant; for example, as depicted in Figure \ref{fig:1.1M}, variations in metallicity alone can significantly alter the stability of systems, even when other conditions remain constant. Secondly, Figure \ref{fig:C1} presents a comparison between our model and the DL15 model regarding the stellar rotation period corresponding to torque balance under solar metallicity conditions.
For systems with solar masses, the differences in trends between the two models are minor, with only numerical variations observed in the calculated $P_{\text{sta}}$ obtained from our model. However, for systems with masses of 1.3 $M_{\odot}$, 1.0 $M_{\text{J}}$. Our model yields smaller $P_{\text{sta}}$ values compared to systems with masses of 1.0 $M_{\odot}$ and 10 $M_{\text{J}}$. In contrast, the DL15 model yields larger $P_{\text{sta}}$ values for systems with masses of 1.3 $M_{\odot}$ and 1.0 $M_{\text{J}}$ compared to those with masses of 1.0 $M_{\odot}$ and 10 $M_{\text{J}}$. This discrepancy arises because, in the DL15 model, the treatment of F stars merely reduces the angular momentum loss rate by an order of magnitude compared to G stars. In contrast, our model employs the \citet{2015ApJ...799L..23M} magnetic braking model, which utilizes the convective turnover timescale to describe variations in stellar mass. Consequently, in the DL15 model, the angular momentum loss due to magnetic braking varies discontinuously with stellar mass, potentially resulting in a reversed trend in the calculated $P_{\text{sta}}$.
For systems with masses of 1.3 $M_{\odot}$ and 10.0 $M_{\text{J}}$, both models yield $P_{\text{sta}}$ values equal to $P_{\text{orb}}$ when $P_{\text{orb}}$ is less than 5.0 days. However, when $P_{\text{orb}}$ exceeds 5 days, the $P_{\text{sta}}$ values obtained by the DL15 model gradually surpass $P_{\text{orb}}$. Finally, the overall trend of our model is consistent with that of the DL15 model, namely, long-term double synchronization tends to occur in systems with larger planetary masses around F stars. However, for real observed systems, the magnetic braking scheme of the DL15 model cannot accurately provide the angular momentum loss rate for each star. Therefore, in the search for long-term double synchronous systems, employing a more sophisticated magnetic braking model such as \citet{2015ApJ...799L..23M} can provide better constraints on stellar parameters.

\onecolumn

\begin{longtable}{p{2cm}lllrrrrrrlll}
\caption{This work considers the stellar and planetary parameters in the considered systems. The full
	table can be found online in a machine readable format.}
\label{tab:tab2}\\

\multicolumn{1}{c}{\centering\textbf{Name}} & \multicolumn{1}{c}{\textbf{$P_{\centering\text{orb}}$}} &
\multicolumn{1}{c}{\centering\textbf{$a$}} & \multicolumn{1}{c}{\textbf{$M_{\centering\text{pl}}$}} &
\multicolumn{1}{c}{\centering\textbf{$T_{\text{eff}}$}} & \multicolumn{1}{c}{\centering\textbf{$R_*$}} &
\multicolumn{1}{c}{\centering\textbf{$M_*$}} & \multicolumn{1}{c}{\centering\textbf{Metallicity}} &
\multicolumn{1}{c}{\centering\textbf{$v\sin i$}} &
\multicolumn{1}{c}{\centering\textbf{$i_{\centering\text{orb}}$}} & \multicolumn{1}{c}{\centering\textbf{$R_{\text{o}}$}} &
\multicolumn{1}{c}{\centering\textbf{$P_{\centering\text{rot}}$}} \\
\multicolumn{1}{c}{\centering\textbf{}} & \multicolumn{1}{c}{\centering\textbf{(days)}} &
\multicolumn{1}{c}{\centering\textbf{(au)}} & \multicolumn{1}{c}{\centering\textbf{($M_J$)}} &
\multicolumn{1}{c}{\centering\textbf{(K)}} & \multicolumn{1}{c}{\centering\textbf{($R_{\odot}$)}} &
\multicolumn{1}{c}{\centering\textbf{($M_{\odot}$)}} & \multicolumn{1}{c}{\centering\textbf{(dex)}} &
\multicolumn{1}{c}{\centering\textbf{($\text{km} \cdot \text{s}^{-1}$)}} &
\multicolumn{1}{c}{\centering\textbf{($\circ$)}} & \multicolumn{1}{c}{\textbf{}} &
\multicolumn{1}{c}{\centering\textbf{(days)}} \\
\hline
\endhead

         WASP-132 &   7.133521 &     0.06700 &   0.41000 &     4775 & 0.74$^{0.02}_{-0.02}$ &     0.80 &   0.220 &          0.9$^{0.8}_{-0.8}$ &           89.6$^{0.3}_{-0.3}$ &       0.422 &       41.585  \\
          WASP-11 &   3.722480 &     0.04375 &   0.49200 &     4900 & 0.77$^{0.04}_{-0.04}$ &     0.81 &   0.120 &         1.04$^{1.2}_{-1.2}$ &          89.03$^{0.5}_{-0.5}$ &       0.381 &       37.442  \\
        HD 189733 &   2.218575 &     0.03100 &   1.12300 &     5040 & 0.76$^{0.01}_{-0.01}$ &     0.81 &  -0.030 &      3.5$^{0.15}_{-0.15}$ &      85.58$^{0.34}_{-0.34}$ &       0.117 &       10.950  \\
         WASP-144 &   2.278315 &     0.03160 &   0.44000 &     5200 & 0.81$^{0.02}_{-0.02}$ &     0.81 &   0.180 &          1.9$^{1.0}_{-1.0}$ &       86.9$^{0.06}_{-0.06}$ &       0.260 &       21.531 \\
          TOI-181 &   4.532000 &     0.05390 &   0.14520 &     4994 & 0.74$^{0.02}_{-0.02}$ &     0.82 &   0.272 &         2.28$^{0.5}_{-0.5}$ &      88.28$^{0.32}_{-0.34}$ &       0.852 &       16.408  \\
            WTS-2 &   1.018707 &     0.01856 &   1.27000 &     5000 & 0.75$^{0.02}_{-0.02}$ &     0.82 &   0.200 &          2.2$^{0.5}_{-0.5}$ &      83.55$^{0.14}_{-0.14}$ &       0.198 &       17.133  \\
            \ldots &   \ldots &     \ldots &   \ldots &     \ldots & \ldots &     \ldots &   \ldots &          \ldots &      \ldots &       \ldots &       \ldots  \\
\end{longtable}

\twocolumn

\bsp	
\label{lastpage}

\end{document}